\documentclass[11pt,a4paper]{article}
\pdfoutput=1
\usepackage{jheppub}

\usepackage{bbding}
\usepackage{pifont}
\usepackage{wasysym}
\usepackage{wrapfig}
\usepackage{amsmath}
\usepackage{verbatim}
\usepackage{amssymb}
\usepackage{inputenc}
\usepackage{textcomp}
\usepackage{appendix}
\usepackage{mathrsfs}
\usepackage{float}
\usepackage{xfrac}
\restylefloat{table}
\newcommand{\e}{\epsilon}

\newcommand{\be}[1]{\begin{equation}\label{#1} }
\newcommand{\ee}{\end{equation}}
\newcommand{\bea}[1]{\begin{eqnarray}\label{#1} }
\newcommand{\eea}{\end{eqnarray}}
\newcommand{\p}{\partial}

\newcommand{\A}{\mathcal{A}}
\newcommand{\B}{\mathcal{B}}

\renewcommand{\L}{\mathcal{L}}
\newcommand{\bL}{\bar{\mathcal{L}}}

\newcommand{\z}{{\bar z}}

\newcommand{\bk}{\bar{k}}

\renewcommand{\a}{\alpha}
\newcommand{\ta}{\tilde{\alpha}}
\newcommand{\C}{\tilde{C}}
\renewcommand{\b}{\beta}
\renewcommand{\t}{\tau}
\newcommand{\g}{\gamma}
\newcommand{\s}{\sigma}

\newcommand{\bes}{\begin{subequations}}
\newcommand{\ees}{\end{subequations}}

\def\cJ{\mathcal{J}}
\def\cP{\mathcal{P}}
\def\cL{\mathcal{L}}
\def\cH{\mathcal{H}}

\def\nn{\nonumber\\}

\def\a{\alpha}
\def\b{\beta}

\def\e{\epsilon}

\def\m{\mu}

\def\be{\begin{equation}}
\def\ee{\end{equation}}
\def\bea#1\ena{\begin{align}#1\end{align}}
\def\pd{\partial}

\title{Boosting to BMS}

\author[a]{Arjun Bagchi,} \author[b]{Aritra Banerjee,} \author[c]{and Hisayoshi Muraki.} \author{\\} %\author[c]{,}

\affiliation[a]{Indian Institute of Technology Kanpur, Kanpur 208016, INDIA.
\\}

\affiliation[b]{Okinawa Institute of Science \& Technology, 1919-1 Tancha, Onna-son, Okinawa 904-0495, JAPAN.\\}

\affiliation[c]{Center for Geometry and Physics, Institute for Basic Science (IBS), Pohang 37673, KOREA.\\}

\emailAdd{abagchi@iitk.ac.in, aritra.banerjee@oist.jp, hmuraki@ibs.re.kr}

\abstract{Bondi-Metzner-Sachs (BMS) symmetries, or equivalently Conformal Carroll symmetries, are intrinsically associated to null manifolds and in two dimensions can be obtained as an In{\"o}n{\"u}-Wigner contraction of the two-dimensional ($2d$) relativistic conformal algebra. Instead of performing contractions, we demonstrate in this paper how this transmutation of symmetries can be achieved by infinite boosts or degenerate linear transformations on coordinates.
Taking explicit cues from the worldsheet theory of null strings, we show boosting the system is equivalent to adding a current-current deformation term to the Hamiltonian. As the strength of this deformation term reaches a critical value, the classical symmetry algebra ``flows" from two copies of Virasoro to the BMS algebra. We further explore the situation where the CFT coordinates are asymmetrically transformed, and degenerate limits lead to chiral theories.}

\begin{document}
\maketitle

\newpage

\section{Introduction}
Null surfaces are ubiquitous in nature, and the physics associated to these have time and again appeared in various circumstances, ranging from Quantum Gravity and black holes to even certain condensed matter systems. These null surfaces compel us to look beyond our comfort zone of pseudo-Riemannian manifolds and known Lorentzian physics. Notable instances of null hypersurfaces include the event horizons of black holes, boundary of causal diamonds and hypersurfaces appearing at light-like infinity of asymptotically flat spacetimes ($\mathscr{I}^{\pm}$). 

%To generalize a bit, a null surface presents itself whenever we encounter a case of geometrical degeneracy. 

%{\cl{AB: This last statement is not true. Think e.g. of Galilean manifolds. These are not null surfaces.}}
\medskip

The Riemannian metric formulation for null surfaces fail since the metric degenerates in this limit and one has to resort to \textit{Carrollian} structures that arise on such surfaces. Carroll group is obtained by a contraction of Poincaré group where the speed of light $c\to 0$, and the associated kinematical structures allow us to define locally Carroll manifolds \cite{LBLL,NDS,Duval:2014uoa,Duval:2014uva,Duval:2014lpa}. These manifolds are endowed with a fibre bundle structure that keeps space and time diffeomorphisms separate from each other. The theories defined on such Carrollian manifolds can also be seen as a $c\to 0$ (often called Ultra-Relativistic) limit of relativistic theories. It is important to note the resultant theories are different from the diametrically opposite Non-Relativistic ($c\to \infty$), or Galilean limits, which find interpretation as the ones defined on a Newton-Cartan manifold \cite{Duval:2014uoa}. However, there are interesting similarities, which become most manifest in the two dimensional theories we would investigate in this paper. 
\medskip

A particular motivation behind the recent resurgence in studies of Carrollian structures stems from interests in asymptotic symmetries of flat spacetimes. It has been long known  from the work of  Bondi, van der Burgh, Metzner and Sachs  \cite{Bondi:1,Sachs:1962zza} that asymptotic symmetry groups on $\mathscr{I}^{\pm}$ have an infinite dimensional structure, and are often called the BMS groups  \cite{Bondi:1,Sachs:1962zza,Barnich:2006av}. Since $\mathscr{I}^{\pm}$ are explicit examples of null manifolds, it is not surprising that the $(d+1)$ dimensional BMS algebra is isomorphic to the $d$ dimensional Carrollian Conformal Algebra (CCA). The structure of BMS symmetries (and hence Carrollian physics) has been instrumental in defining the building blocks of a formulation of the holographic correspondence for asymptotically flat spacetimes where putative dual field theories live on the null boundary which are Carrollian manifolds  \cite{Barnich:2010eb,Bagchi:2010zz,Bagchi:2012cy}. This formulation, more recently called Carroll holography, has met with success in a $3d$ version of flat holography, where the putative dual is a $2d$ BMS invariant conformal field theory (BMSFT) \cite{Bagchi:2012yk,Bagchi:2012xr,Barnich:2012xq,Bagchi:2013qva, Bagchi:2015wna, Bagchi:2014iea,Jiang:2017ecm,Hijano:2017eii, Bagchi:2013lma, Detournay:2014fva, Afshar:2013vka, Gonzalez:2013oaa,Hartong:2015usd, Hartong:2015xda, Bagchi:2016geg, Barnich:2014cwa, Fareghbal:2014qga, Grumiller:2019xna, Ciambelli:2018wre}. In another avenue, called Celestial Holography, holography for asymptotically flat spacetimes formulated as a correspondence between gravity on $4d$ flat space and a $2d$ CFT living on the celestial sphere has found a lot of success in terms of new results in asymptotic symmetries and scattering amplitudes. We refer to the excellent reviews \cite{Strominger:2017zoo, Pasterski:2021rjz, Raclariu:2021zjz} for a detailed exposition of these aspects.  There is a very recently discovered connection between these two formulations \cite{Bagchi:2022emh}, which uses ideas from Celestial holography to show how to connect Carroll CFT correlations naturally to $4d$ scattering amplitudes. For ideas linking the two approaches, see also \cite{Donnay:2022aba}. 
\medskip

In a parallel development in the context of string theory, Carrollian strings have appeared in two distinct, albeit intertwined, situations. These degenerate metric structures may appear on the worldsheet when one considers the tension of the string going to zero, i.e. a \textit{null string theory}. First put forward by Schild \cite{Schild:1976vq}, and later reinvented in \cite{Isberg:1993av}, the idea of null or \textit{tensionless} strings have gathered momentum recently considering the $2d$ null worldsheet is a Carrollian manifold. Consequently the residual gauge symmetry turns out to be governed by the BMS$_3$ algebra. There have been a number of studies associated to aspects of bosonic \cite{Bagchi:2013bga,Bagchi:2015nca, Bagchi:2020fpr}, and supersymmetric \cite{Bagchi:2017cte, Bagchi:2018wsn} tensionless strings which use explicit Carrollian formulation to study peculiarities of such string theories both at classical and quantum levels. One also should note that Carrollian strings can also appear when the target spacetime has some Carrollian structure or an embedded null hypersurface. It has been shown very recently \cite{Bagchi:2020ats, Bagchi:2021ban} that a string worldsheet moving into near horizon spacetime associated to a black hole inherits an induced Carrollian structure and effectively turns tensionless. 
\medskip

Carrollian and conformal Carrollian symmetries have also recently arisen in the context of black hole horizons \cite{Donnay:2019jiz}, cosmology and dark energy \cite{deBoer:2021jej}, and in the study of fractons in condensed matter \cite{Bidussi:2021nmp}. More and more intriguing new avenues governed by this exotic symmetry are being uncovered as one considers interesting corners of relativistically invariant theories. 

\medskip

\begin{figure}[t]\label{lightcone}
\begin {center}
    \includegraphics [scale = 0.47] {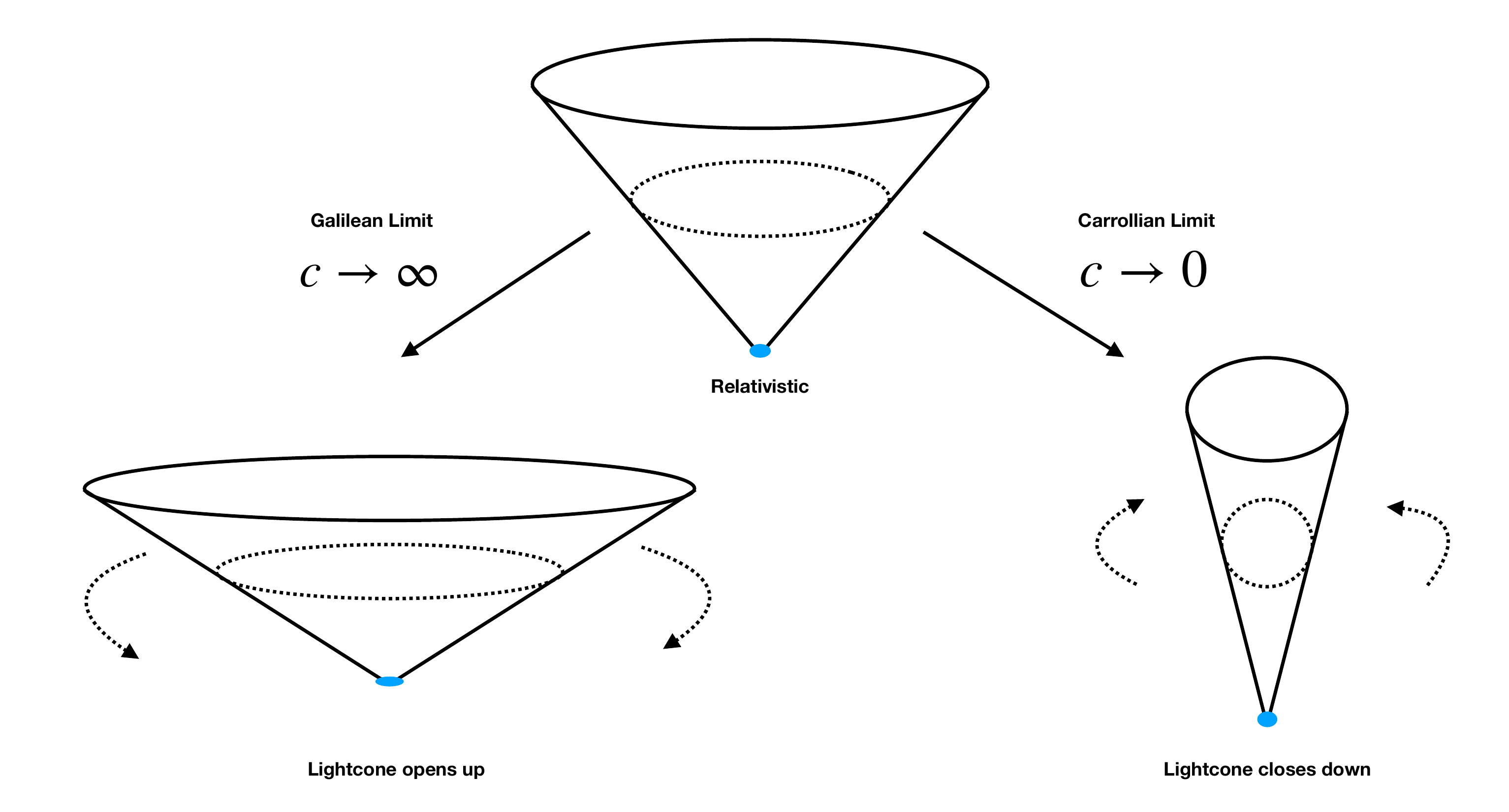} \ \
\caption{Collapsing/Expanding lightcone under Carrollian/Galilean limits.}
\end {center}
\end{figure}

It is thus evident that Conformal Carroll or equivalently BMS algebras are important as symmetry principles in various physical situations. Our present endeavour is to present a more physical perspective on how these symmetries arise. In this work, we will focus on two dimensional theories that are invariant under the BMS$_3$ or equivalently the $2d$ Conformal Carroll algebra. These $2d$ Carrollian Conformal Field Theories (CCFTs) or BMSFTs appear as a $c\to 0$ Inönü-Wigner contraction of $2d$ relativistic CFTs. This points to the fact that lightcones associated to such relativistic theories close up as we take the limit, and effectively become just the time axis in the strict Carrollian limit ($c=0$). One could also think of the opposite Galilean limit ($c\to \infty$) in $2d$, and associated Galilean Conformal Field Theories (GCFTs), where lightcones open up and span the whole physical spacetime in the strict limit, see figure \eqref{lightcone} for an idea. It turns out in both of these cases, the contraction of $2d$ relativistic CFT leads to a classically BMS$_3$ symmetric theory. This is particular to spacetime dimensions $d=2$ where the Galilean and Carrollian conformal algebras are isomorphic \cite{Bagchi:2010zz}. 
\medskip

This brings us to the main question posed in this paper, can we quantify the closing down (or opening up) of the lightcones associated to a relativistic system as a result of \textit{infinite boosts} acting on the theory? In a Lorentzian world, an evolution in boosts should never change any physics. However funny things can happen when these boost becomes infinite. Non-invertible infinite boosts that change the structure of lightcones by singular scalings of the speed of light, can change physics and that too drastically. Here we concentrate on explicitly showing this particular phenomenon, and subsequently reach the conclusion that these particular transformations having degenerate transformation matrices take a relativistic CFT into a BMS invariant field theory. 
\medskip

Focussing on the $2d$ case, and constantly alluding to results from tensionless string theory, we will show how to take a  continuous path to transform CFT (holomorphic-antiholomorphic) quantities into BMS ones by simultaneously boosting and scaling the speed of light via a $SL(2)$ transformation. For a finite transformation, these effects can be undone for a conformal theory, however in the extreme limits where the lightcones close down (or open up), the invertibility is lost, and BMS symmetries naturally emerge at this point. The reader may wonder how our construction here is different from the infinite boost limits studied earlier in literature, e.g. in \cite{Balasubramanian:2009bg} where the authors end up with a chiral half of a $2d$ CFT starting out with a usual $2d$ CFT. Infinite boosts naturally end up putting the original theory on null surfaces and thereby making these theories Carrollian. If one starts with a $2d$ CFT one always would end up with a $2d$ Carroll CFT or equivalently a $2d$ field theory with BMS$_3$ invariance. In certain cases, e.g. \cite{Bagchi:2012yk}, the symmetries of the $2d$ Carroll CFT reduce to just a single copy of the Virasoro algebra. Our discussions and observations are thus not in contradiction to \cite{Balasubramanian:2009bg}. This case is just a special one of null surface symmetries reducing from BMS to its Virasoro subalgebra {\footnote{We thank an anonymous referee for bringing up this point.}}. 
\medskip

Returning to our constructions in this paper, we will show that the ``flow'' from 2d CFT to the theory with BMS symmetries is achieved by adding a current-current bilinear, generated by boost transformations, as a deformation term to the CFT Hamiltonian. This is again reminiscent of what has already been seen in tensionless string literature, and grants a sanity check for our formalism. As a culmination of this, we write down an interpolating algebra that takes one from the conformal algebra to BMS$_3$ via proper dialling of the boost parameter. 
\medskip

To show the usefulness of this simple yet powerful procedure, we further discuss applying asymmetric transformations to (anti)holomorphic quantities which in general have no relation to Lorentz boosts. In a string theoretic setting, it turns out that a subsection of these generic transformations are known in the literature in the garb of an unconventional gauge fixing procedure, known as the Hohm-Siegel-Zwiebach (HSZ) gauge, that replaces the usual conformal gauge. As the idea suggests, the theory loses conformal invariance in this case, but an extreme limit of such a theory also takes one to a null worldsheet reminiscent of an Ambitwistor string theory \cite{Casali:2016atr, Casali:2017zkz}. We use our formulation to show such theories also give rise to BMS$_3$ as the symmetry algebra.
\medskip 

The paper is organized as follows. In section \eqref{null} we revisit the construction of tensionless limits of string worldsheets and consequent emergence of BMS$_3$ as residual symmetry algebras in this limit. We will also elucidate how BMSFT/GCFT appears as contractions of relativistic CFTs. Section \eqref{boosted} is the main part of the paper, where we introduce a formalism to explicitly transform conformal field theories into BMS invariant theories by doing linear transformations that mix left and right sectors of the theory. We will show how this formalism works at every level including that of currents, stress-energy tensors, and equations of motion. We will also write down an interpolating symmetry algebra that connects between the two cases.  In section \eqref{asyboost} we will demonstrate the effect of generic, non-equivalent transformations on left and right sectors of a conformal theory. We will then end with some discussion of our results and a chart of the road ahead. In the appendices \eqref{apa} and \eqref{apb} we will present some details of our transformations and explicit calculations. 

\bigskip

\section{Null Strings and Scalar BMSFT }\label{null}

As mentioned in the introduction, tensionless limit of string theory is one where the string worldsheet becomes a Carrollian (or null) manifold. This can also be identified as the very high-energy limit of conventional string theory \cite{Gross:1}, which shows high temperature phase structures and Hagedorn transitions \cite{Pisarski:1982cn}. For the sake of discussions in this paper, we will have a quick revision of the central concepts of such tensionless strings, and associated symmetry structures.

\subsection{Intrinsic theory}

\noindent Intrinsic discussion of the theory of tensionless strings starts with the ILST action \cite{Isberg:1993av}, 
\be\label{LST}
S_{\text{ILST}} = \int d^2 \s \, V^\a V^\b~ \p_\a X \cdot \p_\b X .
\ee
Here $V^\a$ are certain vector densities, there is an assumed background Minkowski metric where the string propagates and we have suppressed the target space indices. To derive this action, one starts with the Polyakov action for the bosonic tensile strings
\be{}
S_p =-\frac{T_0}{2} \int d^2 \s \, \sqrt{-\g} \g^{\a \b} \p_\a X^\mu \p_\b X^\nu \eta_{\mu \nu},~~~T_0=\frac{1}{2\pi\a'}
\ee
For an explicit theory of tensionless string, the worldsheet metric $\g^{\a \b}$ becomes degenerate and hence has to be replaced by vector densities: $\sqrt{-\g} \g^{\a \b} \to V^\a V^\b$ as $T_0\to 0$. The equations of motion (EOM) of this action $S_{\text{ILST}}$ are given by 
\be\label{eom}
\p_\a(V^\a V^\b \p_\b X^\mu) = 0, \quad V^\b G_{\a \b} =0, 
\ee
where $G_{\a \b}= \p_\a X^\mu \p_\b X^\nu \eta_{\mu \nu}$  is the induced metric on the worldsheet. 

Now remember in the conformal gauge, where we choose $\gamma^{\a\b} = \eta^{\a\b}$, the tensile string action can be written as, 
\be{}\label{conaction}
S_p =-\frac{T_0}{2} \int dzd\z ~\p_z X\p_{\z}X,
\ee
where on the cylinder the (anti)-holomorphic coordinates are, $z,\z = \t\pm \phi$ and the associated (anti)-holomorphic currents are given by $J^\mu = \p_z X^\mu$ and $\bar{J}^\mu = \p_{\z} X^\mu$. In this form, we can use CFT techniques on the worldsheet to map out the symmetries associated to it. It turns out that after fixing the gauge the residual symmetries are generated by the vector fields
\be
\boldsymbol\ell_n = ie^{inz}\p_z,\qquad \bar{\boldsymbol\ell}_n =ie^{in\z}\p_{\z},
\ee
classical constraints corresponding to which generate two copies of the Virasoro (Witt) algebra %classically,
\be
\{\L_n, \L_m\}_{\text{PB}} = -i(n-m) \L_{n+m}, \qquad \{\bL_n, \bL_m\}_{\text{PB}} = -i(n-m)\bL_{n+m}.
\ee
This is also two copies of diffeomorphisms on a circle, i.e. Diff($S^1$)$\times$Diff($S^1$). 
The equivalent of conformal gauge description for the null string action turns out to be the gauge  $V^\alpha = (1, 0)$. The equivalent action simply reads, 
\be
 S_t =\frac{1}{4\pi c'} \int d\t d\phi (\partial_{\t}X)^2. \label{St-cylinder}
\ee
This action is also the starting point of discussion for a BMS$_3$ or Carrollian Conformal invariant scalar field in two dimensions \cite{Hao:2021urq}, when we strip the scalar field $X$ of the target space index. For most of this work we will be interpolating between these two descriptions, but will remind the reader the subtle differences when they arise.
This worldsheet theory enjoys certain diffeomorphism invariance as well, and some gauge symmetry is still left over even after this gauge fixing. Like the case of two copies of the Virasoro symmetries for the  conformal gauge-fixed tensile strings, the residual gauge symmetry algebra here is generated by the following vector fields, 
\be
\ell_n = ie^{in\phi}( \p_\phi+ in\t \p_\t), \qquad m_n =  i  e^{in\phi} \p_\t,
\ee
with the associated symmetry algebra having the well known form: 
%\bea
%[L_n, L_m] &= (n-m) L_{n+m} + c_L\delta_{n+m,0} (n^3-n), \cr
% [L_n, M_m] &= (n-m)M_{n+m} + c_M\delta_{n+m,0} (n^3-n), \cr 
%[M_n, M_m] &=0.
%\ena
\be \label{bms1}
	\{L_n, L_m\}_{\text{PB}} = -i(n-m) L_{n+m},\quad
	\{L_n, M_m\}_{\text{PB}} = -i(n-m) M_{n+m},\quad
	\{M_n, M_m\}_{\text{PB}} = 0.
\ee
This precisely is the classical part of the BMS$_3$ algebra \cite{Bondi:1,Sachs:1962zza}, which as we have stated in the introduction, also arises in the context of gravitational physics as the asymptotic symmetries of $3d$ flat spacetimes at the null boundary. One can immediately note that having this form, the (anti)-holomorphic nature of the algebra is gone in the BMS$_3$ case. This is the main anchoring point of studying strings with a Carrollian worldsheet and has been explored in detail in recent years along many interesting directions \cite{Bagchi:2013bga,Bagchi:2015nca,Bagchi:2016yyf, Bagchi:2017cte, Bagchi:2018wsn, Bagchi:2019cay, Bagchi:2020fpr, Bagchi:2020ats} \footnote{For other related older literature on null strings, the reader can look at \cite{ Karlhede:1986wb, Lizzi:1986nv, Gamboa:1989px, Gustafsson:1994kr, Lindstrom:2003mg}}.

\paragraph{Intrinsic mode expansions:} In the gauge $V^\alpha = (1, 0)$, the EOM for the scalars and the constraints  corresponding to the densities $V^\a$ are:
\be\label{xeom}
\ddot{X}=0,\quad \dot{X}\cdot X'=0, \quad \dot{X}^2=0,
\ee
where dots and primes correspond to derivatives with respect to coordinates on a cylinder, i.e. $\t$ and $\phi$ respectively. We also would like the tensionless worldsheet to maintain the closed string boundary conditions of the form $X^\mu(\tau,\phi)=X^\mu(\tau,\phi+2\pi)$, like the tensile one, then the above EOM can solved by the following mode expansion:
\be\label{mode} 
X^{\mu}(\tau, \phi)=x^{\mu}+\sqrt{\frac{c'}{2}}B^{\mu}_0\tau+i\sqrt{\frac{c'}{2}}\sum_{n\neq0}\frac{1}{n} \left(A^{\mu}_n-in\tau B^{\mu}_n \right)e^{-in\phi}. 
\ee
Here the $c'$ is a finite constant that has been included to take care of the dimensions. $A,B$ are certain oscillator modes and the conjugate momenta is given by $\Pi(\tau,\phi) = \sum_n B_n e^{-in\phi}.$
Using this expansion, it is easy to find the form of constraints generating the symmetry algebra
\be\label{lmab} 
L_n= \frac{1}{2} \sum_{m} A_{- m}\cdot B_{m+n}, 
\qquad
M_n= \frac{1}{2} \sum_{m} B_{-m}\cdot B_{m+n}. 
\ee
Note here $A,B$ are not the harmonic oscillator modes in the usual sense, and using the canonical equal time Poisson brackets for the theory,
\begin{eqnarray}
	&& \{X(\tau,\phi), \Pi(\tau,\phi')\}_{\rm PB} = \delta(\phi-\phi'), \nonumber \\
	&& \{X(\tau,\phi), X(\tau,\phi')\}_{\rm PB}
	 = \{\Pi(\tau,\phi), \Pi(\tau,\phi')\}_{\rm PB} = 0, \label{canonical-Poisson-brackets}
\end{eqnarray}
 we find they satisfy a different poisson bracket structure,
\bea \label{AB1}
\{A_m^\mu,A_n^\nu\}_{\rm PB}= \{B_m^\mu,B_n^\nu\}_{\rm PB}=0;
\qquad
\{A_m^\mu,B_n^\nu\}_{\rm PB}= - 2 im\delta_{m+n}\eta^{\mu\nu}.
\ena
 The algebra of the constraints, of course, still leads to the BMS$_3$ algebra as before. To make things more transparent, we need to transform $(A,B)$ into a harmonic oscillator basis: 
\be\label{CC}
C^{\mu}_n = \frac{1}{2}({A}^{\mu}_n+B^{\mu}_{n}), \quad \C^{\mu}_n =\frac{1}{2}(-{A}^{\mu}_{-n}+B^{\mu}_{-n}).
\ee
It is easy to see that $(C, \C)$ have the usual structure:
\bea \label{AB}
\{C_m^\mu,C_n^\nu\}_{\rm PB}= \{\C_m^\mu,\C_n^\nu\}_{\rm PB}=-  im\delta_{m+n}\eta^{\mu\nu}.
\ena
One can now also write the mode expansions with respect to these oscillators,
%\bea{cexpansion}
%X^\mu(\t,\s)&=&x^\mu+2\sqrt{\frac{c'}{2}}C^\mu_0\t +i\sqrt{\frac{c'}{2}}  \sum_{n\neq 0} \frac{1}{n}\left[(C^\mu_n-\C^\mu_{-n})-in\t (C^\mu_n+\C^\mu_{-n})\right]e^{-in\s} \nonumber
%\eea
%with zero modes given by the momentum,
%\be{zero}
%C^\mu_0=\C^\mu_0=\sqrt{\frac{c'}{2}}k^\mu.
%\ee 
and consequently, the expressions for the classical constraints in this language becomes
\be
\begin{aligned}
\label{LM} 
& L_n=\sum_{m} (C_{-m}\cdot C_{m+n}-\C_{-m}\cdot \C_{m-n})  \\ 
& M_n=\sum_{m} (C_{-m}\cdot C_{m+n}+\C_{-m}\cdot \C_{m-n}+2C_{-m}\cdot \C_{-m-n}).
\end{aligned}
\ee
The form of these generators will be crucial to the later part of this work.

%As we will go on to see, and as explained in detail in earlier work \cite{Bagchi:2020fpr}, the $C$-oscillators would play an important role in the theory of tensionless strings, effectively defining the quantum structures associated to it. 

\paragraph{Manifestly Carrollian Conformal field theories:} Recently studying intrinsically Carrollian CFTs on null manifolds has gained some momentum \cite{Gupta:2020dtl,Henneaux:2021yzg,Hao:2021urq,Chen:2021xkw, deBoer:2021jej, Bagchi:2022eav}. The null string structure we elucidated above falls under a certain class of Carrollian actions. Due to the degenerate nature of the system, under Carroll-diffeomorphisms on two dimensional cylinder $\s^{\mu}\rightarrow \s'^{\mu}(\s^{\nu})$, time and space coordinates change in different manners viz. $\t'=\t'(\t,\phi)$ and $\phi'=\phi'(\phi)$. Specifically for an infinitesimal Carroll Conformal transformation, we will have:
\be\label{supert}
\delta\t = f'(\phi)\t + g(\phi),~\delta\phi = f(\phi), ~~f~ \&~ g~ \text{arbitrary functions.}
\ee
These are the explicit transformations generated by the BMS generators. Carroll boosts, defined on null manifolds, are different from their relativistic counterparts in the sense of space-time asymmetry, and these translate conformal fields on a null manifold. Let us now comment on this point of view for studying null action without going into gory details. 
\medskip

In Carrollian geometry, the Riemannian metric is replaced by two objects: a degenerate metric $h_{\a\b}$ and the no-where vanishing vector field $\tau^{\a}$ that gives the manifold a null time direction, also called a Clock-form \cite{Duval:2014uoa}. The orthogonality between these two ingredients signify $h_{\a\b}\t^\a \t^\b = 0$. Since the metric in Riemannian sense is degenerate, taking contractions of these metric fields $(h,\tau)$ with the derivative of a single scalar field i.e. $\partial_{\a}X$, we can construct different classes of Carroll covariant second order actions. Especially, contracting with $\tau^{\a}$ direction gives rise to the ``timelike" action:
\be\label{taction}
S_t \sim \frac{g}{4\pi} \int d\t d\phi\, e\, \t^{\a}\t^{\b}\,\partial_{\a}X\partial_{\b}X.
\ee
Here $X$ are scalar fields and $g$ is a coupling constant. This action by construction is BMS$_3$ invariant, and can only be defined on a Carrollian manifold. This is the action comparable to our null string action, and clearly we can see the map:
\begin{equation}
	V^{\a} = \sqrt{e} \t^{\a},
\end{equation}
explicitly equates them. We will not be using this geometric language in this work, but it is important to remember that Carrollian theories can be intrinsically defined to reside on null surfaces. For a more detailed discussion on geometric aspects with respect to scalars on $2d$ null manifolds, the reader is referred to \cite{Bagchi:2022eav}. 

\subsection{Limiting analysis}

 As we have seen, intrinsically tensionless strings, or equivalently BMS$_3$ invariant scalar fields are defined with respect to null directions on a two dimensional null manifold. One can also describe these theories by considering a systematic limiting procedure starting from the corresponding relativistic theory. This is essentially a limit where we transform the worldsheet coordinates to an infinitely boosted frame, i.e. the speed of light on the string worldsheet goes to zero $(c\to 0)$. As we have discussed earlier, this particular class of limits are called Carrollian limits and when taken on relativistic CFTs, they lead to corresponding Carrollian CFTs.  In terms of coordinates on the string worldsheet, taking this limit entails \cite{Bagchi:2015nca}
\be\label{URlim}
\phi \to \phi, \qquad \t \to \e \t, \qquad \a'\to c'/\e, \qquad \e \to 0.
\ee
To start with one can see systematically taking this limit on the relativistic action \eqref{conaction} reproduces the null action \eqref{St-cylinder}. This contraction is clearly an Ultra-Relativistic (UR) one where the worldsheet becomes a degenerate manifold and lightcones shrink towards being just along the vertical axis (see figure \eqref{lightcone}). Consequently the Riemannian worldsheet is replaced by a Carroll manifold, which mathematically is a fibre bundle. If one consistently follows through with the above described limit at every stage of computing symmetries, it leads to the following contraction of the Virasoro generators \cite{Barnich:2006av, Bagchi:2012cy}
\be\label{vir2bms}
L_n= \L_n - \bL_{-n}, \qquad M_n = \e(\L_n + \bL_{-n}),
\ee
where these contracted generators again give rise exactly to the classical BMS$_3$ algebra. Note that it is absolutely crucial to take the extreme limit $\e \to 0$ in this case, or for a finite $\e$ the symmetries of the system remains that of the relativistic conformal symmetries as the brackets don't change. Only at the $\e = 0$ the new symmetries emerge out of the Virasoro symmetries. This makes perfect sense since in an evolution in finite relativistic boosts, the physics of a system must not change.

 One can now remember the tensile oscillator mode expansion: 
 \be\label{t-exp} 
X^{\mu}(\phi,\tau)=x^{\mu}+2\sqrt{2\alpha'}\alpha^{\mu}_0\tau+i\sqrt{2\alpha'}\sum_{n\neq0}\frac{1}{n}[\a^{\mu}_ne^{-in(\tau+\phi)}+\ta^{\mu}_ne^{-in(\tau-\phi)}] 
\ee
Comparing with the tensionless expansions under the limit, and a few lines of algebra gives rise to the simple relations between tensile and tensionless oscillators:
 \be\label{ABa}
A_n^{\mu} = \frac{1}{\sqrt{\e}} \left( \a_n^\mu - \tilde{\a}_{-n}^\mu \right),
\qquad 
B_n^{\mu} = {\sqrt{\e}} \left( \a_n^\mu + \tilde{\a}_{-n}^\mu \right). 
\ee
Here $\a_n$ and $\tilde{\a}_n$ are the tensile oscillators coming from the usual mode expansion,  and these annihilate the tensile vacuum, which we can call  $|0\rangle_{\a}$.
All classical physics of the tensionless string can be reproduced by following the UR limit we described above. 
\medskip

Later in the paper, we will also be encountering the exact opposite of the Carrollian limit, in the form of the so-called Non-Relativistic (NR) or Galilean contraction \cite{Bagchi:2009pe}. In this case the $\e$ dependent contraction of the generators turn out to be,
\be\label{NR}
L_n= \L_n + \bL_{n}, \qquad M_n = \e(\L_n -\bL_{n}),
\ee
In two dimensions this again leads to the classical part of the BMS$_3$ algebra starting from two copies of the Virasoro algebra. Written in terms of co-ordinates on the worldsheet, this could easily be envisaged as the very opposite limit i.e. $\phi\to\e\phi, \t\to\t, ~\e\to 0$. This corresponds to the limit where the speed of light on the string worldsheet goes to infinity $(c\to\infty)$.

\bigskip

\section{BMSFT as an infinitely boosted CFT}\label{boosted}
\subsection{Rationale}
We have discussed in the previous section how Carroll contractions of relativistic worldsheet leads to BMS$_3$ invariant theories. In this section, we will see how an infinite boost on our worldsheet theories leads to a clear change in physics. 

\begin{figure}[h!]
\begin {center}
    \includegraphics [scale = 0.17] {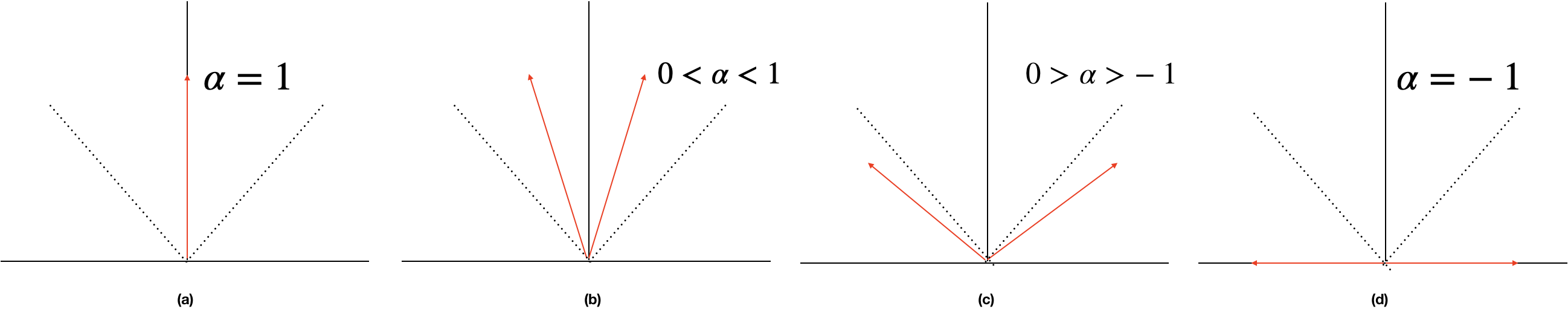} \ \
\caption{A continuous one parameter evolution of lightcone from $\a=1$ (UR Case) to $\a=-1$ (NR Case) in two dimensions. The dotted lines represent $\a=0$, i.e. the relativistic case.}
\label{fig2}
\end {center}
\end{figure}

We start with a special linear ($SL(2)$) transformation of our (anti)-holomorphic coordinates on our (conformal gauge fixed) worldsheet. Let the transformed theory be defined by changed coordinates $z_{L/R}$, which mixes contributions from both holomorphic and anti-holomorphic coordinates:  
\be{}
\label{Symm:z_LR}
	\begin{pmatrix}
	z_{\rm L} \\ z_{\rm R}
	\end{pmatrix}
	=\frac{1}{\sqrt{1-\a^2}}
	\begin{pmatrix}
	1 & -\a \\ -\a & 1
	\end{pmatrix}
	\begin{pmatrix}
	z \\ \overline{z}
	\end{pmatrix}
	\qquad
	\quad
	\ee
For real $\a$ this is a relativistic (quasi-) boost in $(z,\z)$ coordinates. One should remember that on a cylinder $(z,\z)$ are basically lightcone coordinates, which a timelike observer in $(\t,\phi)$ coordinates can never reach via finite Lorentz boosts, hence nature of boost in these coordinates is clearly different. One could actually show that boost transformations along these $(z,\bar{z})$ coordinates are (quasi-) boosts in $(\t,\phi)$ coordinates, albeit up to a rescaling of the speed of light. For more details on this phenomenon, the reader is directed to Appendix \ref{apa}.
This shows up in the transformed two dimensional metric in $(\tau,\phi)$-coordinates as well, which reads:
\be{}\label{dmetric}
\begin{pmatrix}
	\eta^{\tau\tau} & \eta^{\tau\phi} \\ \eta^{\phi\tau} & \eta^{\phi\phi}
	\end{pmatrix}
	=
	\begin{pmatrix}
	\frac{1+\a}{1-\a} & 0 \\
	0 & -\,\frac{1-\a}{1+\a}
	\end{pmatrix}
\ee
So this transformation in these coordinates still keeps the conformal gauge invariant upto some 
rescaling of the coordinates, upon which it becomes the flat Minkowski metric. However this rescaling ceases to work at the singular points $\a = \pm 1$, which again makes perfect sense as at these points the transformation \eqref{Symm:z_LR} stops being invertible, and the metric in \eqref{dmetric} degenerates. Unless one hits these singular point, this set of transformations should not change the worldsheet physics in any shape or form. We can actually see that identifying this transformation as a scaling,
\be
\t \to \,\frac{1-\a}{1+\a}\t = \e\,\tau,
\qquad \phi \to \phi, \qquad T_0 \to \e\,T_0,
\ee
keeps the action \eqref{conaction} invariant, unless $\e = 0$. One can compare the above with the UR contraction \eqref{URlim}, to see that $\e = 0$ point, i.e. BMS point, corresponds to $\a = 1$. Thus it can be expected that when \eqref{Symm:z_LR} stops being revertible, the worldsheet CFT structure is taken over by BMSFT. This will be the main logical point throughout our discussion in this section.
One should also note, the other singular point $\a = -1$ corresponds to $\e \to \infty$, which can be rephrased as the NR contraction of the worldsheet CFT \eqref{NR}.  Reaching these two special points will be our notion of \textit{infinite boosts}, and both will correspond to a change in symmetry algebras. See figure \eqref{fig2} for an illustration of this one parameter evolution between the UR point and NR point. 
%whereas rescaling is equivalent to adjusting the speed of light
%in a similar sense as discussed in Carroll string \rc{[Ref]}.
%As explained in the appendix,
%even though the metric is out of conformal gauge,
%the notion of separation of variables of (anti-)holomorphic sectors still applies.

\subsection{Virasoro generators and Sugawara construction}
Let us first set our CFT notations straight for the rest of this paper. Since the starting point of our theory is a $2d$ CFT, we assume that there exists a pair of stress-energy tensors subject to the following %fundamental 
Poisson brackets, which give the classical conformal algebra:
\be
\begin{aligned}\label{conformal}
	\{T(\phi), T(\phi')\}_{\rm PB}
	&=2T(\phi)\pd_\phi\delta(\phi-\phi')
	+\pd_\phi T(\phi)\delta(\phi-\phi')
	\\
	\{\overline{T}(\phi), \overline{T}(\phi')\}_{\rm PB}
	&=-2\overline{T}(\phi)\pd_\phi\delta(\phi-\phi')
	-\pd_\phi \overline{T}(\phi)\delta(\phi-\phi')
	\\
	\{{T}(\phi), \overline{T}(\phi')\}_{\rm PB}
	&=0.
\end{aligned}
\ee
In the quantum case, the Poisson brackets are as usual replaced with %Dirac brackets. 
commutators.
These stress tensors come with Fourier expansions in terms of Virasoro generators,
\be
	T(\phi)=\sum_m \L_m\,e^{-im\phi}
	\qquad\quad
	\overline{T}(\phi)=\sum_m \overline{\L}_m\,e^{+im\phi},
\ee
where the $\L$'s generate two disjoint copies of the (classical part of) Virasoro algebra.  
Note that the algebra of anti-holomorphic stress tensors do not change under the transformation $\overline{T}(\phi)\to -\overline{T}(-\phi)$, which enforces the following automorphism of the Virasoro algebra $ \overline{\L}_m\to - \overline{\L}_{-m}$. \footnote{This automorphism is in the heart of a duality map between UR and NR contractions of the Virasoro algebra, shown in \eqref{vir2bms} and \eqref{NR}.  }
The current algebra (affine-Lie algebra) for the theory is given by a set of fields
\be
	J^a(\phi)=\sum_m J_m^a e^{-im\phi}
\ee
%usually the currents are 
which in Dirac bracket sense is equipped with the affine Kac-Moody algebraic relations
\be\label{AffKMrel}
	[J^a_m,J^b_n]=if^{ab}_cJ^c_{m+n}+\frac{k}{2}m\delta^{ab}\delta_{m+n}.
\ee
Here $f^{ab}_c$ are structure constants 
with totally anti-symmetric indices $f^{ab}_c=f^{bc}_a=-f^{ba}_c$,
and
$k$ is the level. The algebra can also be represented by the Poisson structure defined by:
\be
	\{J^a(\phi), J^b(\phi')\}_{\rm PB}
	=\left(f^{ab}_c J^c(\phi)+\frac{k}{2}\delta^{ab}\pd_\phi\right)\delta(\phi-\phi').
\ee
This is true because, on one hand
\be
	\{J^a(\phi), J^b(\phi')\}_{\rm PB}
	=\sum_{m,n}
	\{J^a_m, J^b_n\}_{\rm PB}\,e^{-i(m\phi+n\phi')}
\ee
on the other hand we can see,
\be
	\left(f^{ab}_c J^c(\phi)+\frac{k}{2}\delta^{ab}\pd_\phi\right)\delta(\phi-\phi')
	=\sum_{m,n}
	\left(f^{ab}_cJ^c_{m+n}-i\frac{k}{2}m\delta^{ab}\delta_{n,-m}\right)e^{-i(m\phi+n\phi')}
\ee
so that the algebra of the modes turn out to be
\be
	\{J^a_m, J^b_n\}_{\rm PB}=f^{ab}_cJ^c_{m+n}-i\frac{k}{2}m\delta^{ab}\delta_{m+n},
\ee
which is the classical version of \eqref{AffKMrel}.
%the algebra under the replacement $\{\,\cdot\,,\,\cdot\,\}_{\rm PB}\to -i[\,\cdot\,,\,\cdot\,]$.
For later purpose, note that for the anti-holomorphic current, the Fourier modes are expanded in a form
\be
	\overline{J}^a(\phi)=\sum_m \overline{J}_m^a e^{+im\phi}.
\ee
The Poisson bracket for these currents is conventionally chosen to be
\be
	\{\overline{J}^a(\phi), \overline{J}^b(\phi')\}_{\rm PB}
	=\left(\bar{f}^{ab}_c \overline{J}^c(\phi)-\frac{\bar{k}}{2}\delta^{ab}\pd_\phi\right)\delta(\phi-\phi')
\ee
so as to give
\be
	\{\overline{J}^a_m, \overline{J}^b_n\}_{\rm PB}
	=\bar{f}^{ab}_c\overline{J}^c_{m+n}-i\frac{\bar{k}}{2}m\delta^{ab}\delta_{m+n}.
\ee
The Virasoro generators are obtained as the Fourier components of a bilinear form of the currents known as the Sugawara construction. In the (classical) regime considered here, inspired by the Sugawara construction,
let us introduce the following fields
\be\label{suga}
	T(\phi)=\frac{1}{k}\sum_a J^a(\phi)J^a(\phi),
	\qquad
	\overline{T}(\phi)=\frac{1}{{\bk}}\sum_{{a}} \overline{J}^{{a}}(\phi)\overline{J}^{{a}}(\phi)
\ee
which are shown to obey the relations, in addition to satisfying \eqref{conformal},
\bea\label{TJ}
	\{T(\phi),J^a(\phi')\}_{\rm PB}
	&=J^a(\phi)\pd_\phi\delta(\phi-\phi')+\pd_\phi J^a(\phi)\delta(\phi-\phi')
	\\
	\{\overline{T}(\phi),\overline{J}^a(\phi')\}_{\rm PB}
	&=-\overline{J}^a(\phi)\pd_\phi\delta(\phi-\phi')
	-\pd_\phi \overline{J}^a(\phi)\delta(\phi-\phi').
%	\\
%	\{T(\phi),T(\phi')\}_{\rm PB}
%	&=2T(\phi)\pd_\phi\delta(\phi-\phi')
%	+\pd_\phi T(\phi)\delta(\phi-\phi')
%	\\
%	\{\overline{T}(\phi),\overline{T}(\phi')\}_{\rm PB}
%	&=-2\overline{T}(\phi)\pd_\phi\delta(\phi-\phi')
%	-\pd_\phi \overline{T}(\phi)\delta(\phi-\phi')
\ena
These equations give nothing but a representation of
the Virasoro algebra, %by introducing $T(\phi)=\sum_m \L_m \,e^{-im\phi}$, $\overline{T}(\phi)=\sum_m \overline{\L}_m\,e^{+im\phi}$,
and they confirm that the fields $J$  transform like  
a field with conformal weight one, i.e. a current:
\be\label{}
	\{\L_m,J_n^a\}_{\rm PB} = inJ^a_{m+n},
	\qquad
	\{\bL_m,\bar{J}_n^a\}_{\rm PB} = in\bar{J}^a_{m+n}.
\ee
 For details of the calculation of these brackets, the reader can refer to Appendix \ref{apb}. 
\medskip

For sake of completeness, let us explain how to write the direct product algebras corresponding to 
holomorphic and anti-holomorphic sectors in a semi-direct basis. 
Let us introduce
a pair of combinations of currents weighted by factors of following form,
\be
	\mathbf{J}^a %= d_{\mathbf{J}}(J_L^a -J_R^a)
	=c_{\mathbf{J}}(J^a-\overline{J}^a),
	%\qquad \left(c_{\mathbf{J}}=\sqrt{\frac{1-\a}{1+\a}}d_{\mathbf{J}}\right)\\
	\qquad
	\mathbf{P}^a %= d_{\mathbf{P}}(J_L^a +J_R^a)
	=c_{\mathbf{P}}(J^a+\overline{J}^a).
	%\qquad \left(c_{\mathbf{P}}=\sqrt{\frac{1+\a}{1-\a}}d_{\mathbf{P}}\right)
\ee
Here $c_{\mathbf{J}}$ and $c_{\mathbf{P}}$
are constants, and a simple calculation yields the Poisson brackets,
\be
\begin{aligned}
	\{\mathbf{J}^a(\phi), \mathbf{J}^b(\phi')\}_{\rm PB}
	&=c_{\mathbf{J}}^2\left(f^{ab}_c {J}^c(\phi)+\bar{f}^{ab}_c \overline{J}^c(\phi)
	    +\frac{k_1}{2}\delta^{ab}\pd_\phi\right)\delta(\phi-\phi'),
	\\
	\{\mathbf{J}^a(\phi), \mathbf{P}^b(\phi')\}_{\rm PB}
	&=c_{\mathbf{J}}c_{\mathbf{P}}\left(f^{ab}_c {J}^c(\phi)-\bar{f}^{ab}_c \overline{J}^c(\phi)
	    +\frac{k_2}{2}\delta^{ab}\pd_\phi\right)\delta(\phi-\phi'),
	\\
	\{\mathbf{P}^a(\phi), \mathbf{P}^b(\phi')\}_{\rm PB}
	&=c_{\mathbf{P}}^2\left(f^{ab}_c {J}^c(\phi)+\bar{f}^{ab}_c \overline{J}^c(\phi)
	    +\frac{k_1}{2}\delta^{ab}\pd_\phi\right)\delta(\phi-\phi').		
\end{aligned}
\ee
with the modified levels $k_{1} = k - \bar{k}$ and $k_{2} = k + \bar{k}$.
Making the choice of $\bar{f}^{ab}_c=-{f}^{ab}_c$ makes sure the algebra closes to a current algebra in this basis, upto the rescalings $\mathbf{J}\to \mathbf{J}/c_{\mathbf{J}}$ and $\mathbf{P}\to \mathbf{P}/c_{\mathbf{P}}$. Further assuming the identification  $k=\bar{k}$, 
which is the case of interest for the present paper,
the algebra above is reduced to:
\be\label{JPcurrentAlg}
\begin{aligned}
	\{\mathbf{J}^a(\phi), \mathbf{J}^b(\phi')\}_{\rm PB}
	&=c_{\mathbf{J}}f^{ab}_c \mathbf{J}^c(\phi)\delta(\phi-\phi'),
	\\
	\{\mathbf{J}^a(\phi), \mathbf{P}^b(\phi')\}_{\rm PB}
	&=c_{\mathbf{J}}\left(f^{ab}_c \mathbf{P}^c(\phi)
	    +c_{\mathbf{P}}\,k \delta^{ab}\pd_\phi\right)\delta(\phi-\phi'),
	\\
	\{\mathbf{P}^a(\phi), \mathbf{P}^b(\phi')\}_{\rm PB}
	&=\frac{c_{\mathbf{P}}^2}{c_{\mathbf{J}}} 
	f^{ab}_c \mathbf{J}^c(\phi)\delta(\phi-\phi').		
\end{aligned}
\ee
In particular when the group structure is $U(1)$, i.e. all the structure constants ${f}^{ab}_c$ are zero,
the algebra coincides with \eqref{AB1} when we identify $A,B$ to be the Fourier modes of the currents $\mathbf{J}$ and $\mathbf{P}$, as evidently only the $\{\mathbf{J}, \mathbf{P}\}$ survives. In the limiting sense, referring to the definitions in \eqref{ABa},
we can identify the fourier modes of $J$ and $\overline{J}$ as the tensile oscillators $\a$ and $\tilde{\a}$, with 
$c_{\mathbf{J}}=1/\sqrt{\e}$ and $c_{\mathbf{P}}=\sqrt{\e}$. Note that for $U(1)$ current algebra, any value of $\e$ gives rise to the same bracket structures as above, as long as ${f}^{ab}_c = 0$.
\medskip

Armed with these definitions, we also note that there is a rephrasing of the algebra for a set of stress-energy tensors 
in a different (semi-direct) way:
\be
\label{AlgUndeformJP}
\begin{aligned}
	\{\cJ(\phi),\cJ(\phi')\}_{\rm PB}
	&=2\cJ(\phi)\pd_\phi\delta(\phi-\phi')
	+\pd_\phi \cJ(\phi)\delta(\phi-\phi')
	\\
	\{\cJ(\phi),\cP(\phi')\}_{\rm PB}
	&=2\cP(\phi)\pd_\phi\delta(\phi-\phi')
	+\pd_\phi \cP(\phi)\delta(\phi-\phi')
	\\
	\{\cP(\phi),\cP(\phi')\}_{\rm PB}
	&=2\cJ(\phi)\pd_\phi\delta(\phi-\phi')
	+\pd_\phi \cJ(\phi)\delta(\phi-\phi')
\end{aligned}
\ee
where we define 
\be
\label{UndeformJP}
\begin{aligned}
	\cJ(\phi)&=T(\phi)-\overline{T}(\phi)
	\overset{k=\bar{k}}{=}
	\frac{1}{k}\sum_a \left\{J^a(\phi)J^a(\phi) - \overline{J}^{{a}}(\phi)\overline{J}^{{a}}(\phi)\right\}
	\\
	\cP(\phi)&=T(\phi)+\overline{T}(\phi)
	\overset{k=\bar{k}}{=}
	\frac{1}{k}\sum_a \left\{J^a(\phi)J^a(\phi) + \overline{J}^{{a}}(\phi)\overline{J}^{{a}}(\phi)\right\}
\end{aligned}
\ee
This rephrasing will make it easy to get an interpolation from Virasoro algebra to BMS algebra as described in the next subsection.

\subsection{Flowing from CFT to BMS}

We now move on to the most crucial point of the paper: appearance of BMS from degenerate transformations on CFT. 
In the last section we discussed a manifest semi-direct product structure of Virasoro-Kac-Moody algebras. The current algebra case for $U(1)$   \eqref{JPcurrentAlg} turned out to be straightforward to understand and immediately applicable to our case in \eqref{ABa}, where  $\{\mathbf{P}^a,\mathbf{P}^b\}_{\rm PB}=0$   is the hallmark of appearance of the BMS-Kac-Moody algebra. A non abelian generalization of this turns out to be rather tricky with all structure constants present. In what follows, we will be mostly talking about generic non-Abelian currents, but switching to the Abelian case when we make contact will null string theory. 
%\rc{[ref:1707.07209]}.
%However, in the case of our interest here, the method is no longer applied because of the absence of 
%ways to decouple $\{\mathbf{P}^a,\mathbf{P}^b\}_{\rm PB}$ from \eqref{JPcurrentAlg} in a
%consistent fashion. 

%In other words, a direct product structure of currents takes place instead of a semi-direct product.
%Our construction presented in the following is rather a non-abelian generalization
%along the way of \eqref{LM}.

\subsubsection*{Currents and energy-momentum tensors}

Now we will be interested to take CFT relations written in (anti)-holomorphic coordinates and manifestly boost them to the degenerate limit we earlier talked about.
To start with, under the linear transform \eqref{Symm:z_LR}, 
the corresponding currents %$(J\sim \p X)$ 
shall be transformed in the following way,
\be\label{currentchange}
	\begin{pmatrix}
	J^a_{\rm L} \\ J^a_{\rm R}
	\end{pmatrix}
	=\frac{1}{\sqrt{1-\a^2}}
	\begin{pmatrix}
	1 & \a \\ \a & 1
	\end{pmatrix}
	\begin{pmatrix}
	J^a(z) \\ \overline{J}^a(\bar{z})
	\end{pmatrix}
\ee
owing to the chain rule, e.g. 
$\pd_{\rm L}=\left(\pd z/\pd_{\rm L}\right)\pd_z+\left(\pd \bar{z}/\pd_{\rm L}\right)\pd_{\bar{z}}$. 
%However as long as this transformation is invertible, one can still find a combination that satisfies a (modified) Virasoro-Kac-Moody algebra. 
%The energy-momentum tensor for general values of $\a$ will similarly be modified for this case, taking the form,
%where the generators $\cP$ and $\cJ$ can be expressed as a combination of CFT stress-energy tensors $T, \overline{T}$ and a bilinear of the currents $J, \overline{J}$:
Now, under this transformation, \eqref{UndeformJP} correspondingly gets modified as,
\be
\begin{aligned}\label{bmsff}
	\cJ(\a;\phi)
	&=\frac{1-\a^2}{1+\a^2}\cdot\frac{1}{k}\sum_a \left(J^a_{\rm L}(\phi)J^a_{\rm L}(\phi)-J^a_{\rm R}(\phi)J^a_{\rm R}(\phi)\right) =T(\phi)-\overline{T}(\phi)\\
	\cP(\a;\phi)
	&=\frac{1-\a^2}{1+\a^2}\cdot
	\frac{1}{k}\sum_a \left(J^a_{\rm L}(\phi)J^a_{\rm L}(\phi)+J^a_{\rm R}(\phi)J^a_{\rm R}(\phi)\right)\\
	&=T(\phi)+\overline{T}(\phi)
	+\frac{2\widetilde{\a}}{k}\sum J^a(\phi)\overline{J}^a(\phi)
	~~
	\left(\text{where}~\widetilde{\a}=\frac{2\a}{1+\a^2}\right)
\end{aligned}
\ee
Written in this form, these fields give rise to the algebra similar to \eqref{AlgUndeformJP}:
\be\label{bms.al}
\begin{aligned}
	\{\cJ(\a;\phi),\cJ(\a;\phi')\}_{\rm PB}
	&=\{\partial_{\phi},\cJ(\a;\phi)\}\delta(\phi-\phi')
	\\
	\{\cJ(\a;\phi),\cP(\a;\phi')\}_{\rm PB}
	&=\{\partial_{\phi},\cP(\a;\phi)\}\delta(\phi-\phi')
	\\
	\{\cP(\a;\phi),\cP(\a;\phi')\}_{\rm PB}
	&=(1-\widetilde{\a}^2)\{\partial_{\phi},\cJ(\a;\phi)\}\delta(\phi-\phi').
\end{aligned}
\ee
In the above, we use the shorthand notations $\{\p_\phi,\cJ\} = 2\cJ\p_\phi+ \p_\phi\cJ$ and a similar one for $\cP$. For details of the calculation of these brackets, one can refer to Appendix  \ref{appc}.
From this algebra it is again evident as long as $\widetilde{\a}\neq \pm1$, 
equivalently ${\a}\neq \pm1$,
we can always write a rescaled version of $\cP$ so that we get back to \eqref{AlgUndeformJP},
or equivalently the conformal algebra \eqref{conformal} via boosted stress tensors
\be
\label{EM_LR}
\begin{aligned}
	{T}_{\rm L}(\a;\phi)
	&=\frac12\left(\frac{1}{\sqrt{1-\widetilde{\a}^2}}{\cP}(\a;\phi)+{\cJ}(\a;\phi)\right)\\
	{T}_{\rm R}(\a;\phi)
	&=\frac12\left(\frac{1}{\sqrt{1-\widetilde{\a}^2}}{\cP}(\a;\phi)-{\cJ}(\a;\phi)\right).
\end{aligned}
\ee
 This rescaling stops making sense precisely at $\a=\pm1$ and the above algebra can immediately be identified with the classical BMS$_3$ algebra at that point. Hence, we can continuously connect the Conformal and BMS algebras via our boost transformations parameterised by $\a$. 

Now remember, In terms of the Fourier modes,  the currents are written as $J^a(\phi)=\sum_m J_m^a e^{-im\phi}$ and
$\overline{J}^a(\phi)=\sum_m \overline{J}_m^a e^{+im\phi}$. Using this and the expressions in \eqref{bmsff}, we can write the stress tensor modes in terms of oscillator Fourier modes,
\bea
	\cJ(\a;\phi)
	&=\sum_n\cJ_n(\a)\,e^{-in\phi}
	\qquad
	\cJ_n(\a)
	=\frac{1}{k}\sum_{a,m}  \left(J_{-m}^aJ_{m+n}^a-\overline{J}_{-m}^a\overline{J}_{m-n}^a\right)
	\\
	\cP(\a;\phi)
	&=\sum_n\cP_n(\a)\,e^{-in\phi}
	\qquad
	\cP_n(\a)
	=\frac{1}{k}\sum_m  \left(J_{-m}^aJ_{m+n}^a+\overline{J}_{-m}^a\overline{J}_{m-n}^a
	+2\widetilde{\a} J_{-m}^a\overline{J}_{-m-n}^a\right).
\ena

For a bosonic string, boost dependent currents are related to the interpolating oscillators $C,\C$ \eqref{CC}. Thus, we can take
$J^\mu(\phi)=\pd X^\mu=\sum_m C_m^\mu e^{-im\phi}$ and 
$\overline{J}^\mu(\phi)=\overline{\pd} X^\mu=\sum_m\tilde{C}_m^\mu e^{+im\phi}$ 
with $f^{\mu\nu}_\rho=0$ and $k=\overline{k}=2$. Then the above expressions, at $\a=1$, perfectly match with what we computed from an intrinsic null string theory in \eqref{LM}, i.e. $\cJ_n(\a=1)=L_n$ and $\cP_n(\a=1)=M_n$. In this sense, the transformation from $\a=0$ to $\a=1$ takes the associated string theory from a tensile to a tensionless one. Consequently, the residual symmetry algebra ``flows'' from two copies of Virasoro to the BMS algebra. 
\medskip

Since the basics of the construction are now clear, let us try to alternately define the set of fields $\cJ$ and $\cP$ by introducing a particular nonlinear class of deformation:
\begin{subequations}
\begin{align}
	\cJ(\phi)&=T(\phi)-\overline{T}(\phi)\\
	\cP(\phi)&=T(\phi)+\overline{T}(\phi)+2\tilde{\a}(T(\phi)\overline{T}(\phi))^\b.
\end{align}
\end{subequations}
One can now compare these with the ones we derived in \eqref{bmsff}.
It can be easily seen they satisfy the classical BMS$_3$ algebra when we put the values $\tilde{\a}=1, \b=1/2$,
%(Although it can be merely a consequence of dimensional analysis,
%algebra requires it is the case as discussed below)
\begin{subequations}
\bea
	\{\cJ(\phi),\cJ(\phi')\}_{\rm PB}
	&=\{\partial_{\phi},\cJ(\phi)\}\delta(\phi-\phi')
	\\
	\{\cJ(\phi),\cP(\phi')\}_{\rm PB}
	&=\{\partial_{\phi},\cP(\phi)\}\delta(\phi-\phi')
	\\
	\{\cP(\phi),\cP(\phi')\}_{\rm PB}
	&=0.
\ena
\end{subequations}
Curiously, this is the case defined in \cite{Rodriguez:2021tcz}, where the authors argue that a $\sqrt{T\overline{T}}$ type marginal deformation to the CFT hamiltonian deforms the conformal symmetries into that of BMS. These classical $\sqrt{T\overline{T}}$ deformations are different than their well known irrelevant cousins given by $T\overline{T}$ deformations\footnote{See \cite{Jiang:2019epa} (and references therein) for a very well-crafted introduction to such irrelevant deformations.}, where the $T, \overline{T}$ explicitly correspond to the stress tensors of the deformed theory and the energy of the system is determined by an iterative flow equation. 

For our case, using the definitions in \eqref{suga} for current algebra construction, it is clear that the $\sqrt{T\overline{T}}$-term is explicitly given by the current-current interaction:
\be
	\sqrt{T(\phi)\overline{T}(\phi)}
	\ \simeq\ \frac{1}{k}\sum_aJ^a(\phi)\overline{J}^a(\phi),
\ee
thereby establishing an equivalence of our boost-driven formalism to that in \cite{Rodriguez:2021tcz}. Certainly at the level of Poisson brackets, this equivalence holds well and good. In \cite{Rodriguez:2021tcz}, the authors define this term as an ad-hoc deformation, which has to be added by hand to jump into BMS symmetries \footnote{Also note that in a quantum setting, a square root operator may turn out to be ill-defined.}. We, on the other hand, can physically interpret this term as an infinite limit of finite boost transformations on the CFT, and as we have shown, the strength of the term induces a flow throughout which the theory remains conformal, and at the very end the Carrollian symmetries emerge.

\subsubsection*{Current algebras}

As a  further remark, let us point out the algebra concerning following (normalized) currents
\be
\begin{aligned}
	\mathbf{J}^a(\a;\phi)
	&=\sqrt{\frac{1+\a}{1-\a}}(J^a_L-{J}^a_R)(\phi)
	=J^a(\phi)-\overline{J}^a(\phi)
	\ \left(=\mathbf{J}^a(0;\phi)\right)
	\\
	\mathbf{P}^a(\a;\phi )
	&=\sqrt{\frac{1-\a}{1+\a}}(J^a_L+{J}^a_R)(\phi)
	=J^a(\phi)+\overline{J}^a(\phi)
	\ \left(=\mathbf{P}^a(0;\phi)\right).
\end{aligned}
\ee
Note that they do not explicitly depend on the deformation parameter $\a$ and thus
hereafter we would omit the argument $\a$.
As one can easily deduce, the algebra satisfied by them is \eqref{JPcurrentAlg} with $c_{\mathbf{J}}=c_{\mathbf{P}}=1$.
%\be
%\begin{aligned}
%	\{\mathbf{J}^a(\phi), \mathbf{J}^b(\phi')\}_{\rm PB}
%	&=f^{ab}_c \mathbf{J}^c(\phi)\delta(\phi-\phi'),
%	\\
%	\{\mathbf{J}^a(\phi), \mathbf{P}^b(\phi')\}_{\rm PB}
%	&=\left(f^{ab}_c \mathbf{P}^c(\phi)
%	    +k \delta^{ab}\pd_\phi\right)\delta(\phi-\phi'),
%	\\
%	\{\mathbf{P}^a(\phi), \mathbf{P}^b(\phi')\}_{\rm PB}
%	&=f^{ab}_c \mathbf{J}^c(\phi)\delta(\phi-\phi').		
%\end{aligned}
%\ee
Some straightforward manipulations show the current algebra with respect to $\cJ(\a;\phi)$ and $\cP(\a;\phi)$ has the form,
\be
	\{\cJ(\a;\phi), \mathbf{J}^a(\phi')\}_{\rm PB}
%	&=\{T(\phi)-\overline{T}(\phi),J^a(\phi')-\overline{J}^a(\phi')\}_{\rm PB}\nn
%	&=\{T(\phi),J^a(\phi')\}_{\rm PB}
%		+\{\overline{T}(\phi),\overline{J}^a(\phi')\}_{\rm PB}\nn
%	&=\left(J^a(\phi)-\overline{J}^a(\phi)\right)\pd_\phi\delta(\phi-\phi')
%	+\pd_\phi \left(J^a(\phi)-\overline{J}^a(\phi)\right)\delta(\phi-\phi')\nn
%	&
	=\pd_\phi\left[\mathbf{J}^a(\phi)\delta(\phi-\phi')\right]
\ee
\be
	\{\cJ(\a;\phi), \mathbf{P}^a(\phi')\}_{\rm PB}
%	&=\{T(\phi)-\overline{T}(\phi),J^a(\phi')+\overline{J}^a(\phi')\}_{\rm PB}\nn
%	&=\{T(\phi),J^a(\phi')\}_{\rm PB}
%		-\{\overline{T}(\phi),\overline{J}^a(\phi')\}_{\rm PB}\nn
%	&
	=\pd_\phi\left[\mathbf{P}^a(\phi)\delta(\phi-\phi')\right]
\ee
while
\be
	\{\cP(\a;\phi), \mathbf{J}^a(\phi')\}_{\rm PB}
%	&=\{T(\phi)+\overline{T}(\phi),
%		J^a(\phi')-\overline{J}^a(\phi')\}_{\rm PB}
%		+\frac{2\widetilde{\a}}{k}\sum_b \{J^b(\phi)\overline{J}^b(\phi),
%		J^a(\phi')-\overline{J}^a(\phi')\}_{\rm PB}\nn
%	&=\{T(\phi),J^a(\phi')\}_{\rm PB}
%		-\{\overline{T}(\phi),\overline{J}^a(\phi')\}_{\rm PB}\nn&\qquad
%		+\frac{2\widetilde{\a}}{k}\sum_b \left(
%		\{J^b(\phi),J^a(\phi')\}_{\rm PB}\overline{J}^b(\phi)
%		-\{\overline{J}^b(\phi),\overline{J}^a(\phi')\}_{\rm PB}J^b(\phi)\right)\nn
%	&=\pd_\phi\left[\mathbf{P}^a(\phi)\delta(\phi-\phi')\right]\nn&\qquad
%		+\frac{2\widetilde{\a}}{k}\sum_{b} \left\{
%		\left(\sum_{c}f^{ba}_cJ^c(\phi)+\frac{k}{2}\delta^{ba}\pd_{\phi}\right)
%			\delta(\phi-\phi')\overline{J}^b(\phi)\right\}\nn&\qquad
%		-\frac{2\widetilde{\a}}{k}\sum_{b} \left\{
%		\left(-\sum_{c}f^{ba}_c\overline{J}^c(\phi)-\frac{k}{2}\delta^{ba}\pd_{\phi}\right)
%			\delta(\phi-\phi')J^b(\phi)\right\}\nn
%	&=\pd_\phi\left[\mathbf{P}^a(\phi)\delta(\phi-\phi')\right]
%		+\widetilde{\a}\pd_{\phi}
%		\left[\left(J^a(\phi)+\overline{J}^a(\phi)\right)\delta(\phi-\phi')\right]\nn&\qquad
%		+\frac{2\widetilde{\a}}{k}\sum_{b,c} \left\{
%		f^{ba}_c\left( J^c(\phi)\overline{J}^b(\phi)+\overline{J}^c(\phi)J^b(\phi)\right)
%		\delta(\phi-\phi')\right\}\nn
%	&
	=(1+\widetilde{\a})\pd_\phi\left[\mathbf{P}^a(\phi)\delta(\phi-\phi')\right]
\ee
and most importantly,
\be\label{calPbfP.alg}
	\{\cP(\a;\phi), \mathbf{P}^a(\phi')\}_{\rm PB}
%	&=\{T(\phi)+\overline{T}(\phi),
%		J^a(\phi')+\overline{J}^a(\phi')\}_{\rm PB}
%		+\frac{2\widetilde{\a}}{k}\sum_b \{J^b(\phi)\overline{J}^b(\phi),
%		J^a(\phi')+\overline{J}^a(\phi')\}_{\rm PB}\nn
%	&=\{T(\phi),J^a(\phi')\}_{\rm PB}
%		+\{\overline{T}(\phi),\overline{J}^a(\phi')\}_{\rm PB}\nn&\qquad
%		+\frac{2\widetilde{\a}}{k}\sum_b \left(
%		\{J^b(\phi),J^a(\phi')\}_{\rm PB}\overline{J}^b(\phi)
%		+\{\overline{J}^b(\phi),\overline{J}^a(\phi')\}_{\rm PB}J^b(\phi)\right)\nn
%	&=\pd_\phi\left[\mathbf{J}^a(\phi)\delta(\phi-\phi')\right]\nn&\qquad
%		+\frac{2\widetilde{\a}}{k}\sum_{b} \left\{
%		\left(\sum_{c}f^{ba}_cJ^c(\phi)+\frac{k}{2}\delta^{ba}\pd_{\phi}\right)
%			\delta(\phi-\phi')\overline{J}^b(\phi)\right\}\nn&\qquad
%		+\frac{2\widetilde{\a}}{k}\sum_{b} \left\{
%		\left(-\sum_{c}f^{ba}_c\overline{J}^c(\phi)-\frac{k}{2}\delta^{ba}\pd_{\phi}\right)
%			\delta(\phi-\phi')J^b(\phi)\right\}\nn
%	&=\pd_\phi\left[\mathbf{J}^a(\phi)\delta(\phi-\phi')\right]
%		-\widetilde{\a}\pd_{\phi}
%		\left[\delta(\phi-\phi')\left(J^a(\phi)-\overline{J}^a(\phi)\right)\right]\nn&\qquad
%		+\frac{2\widetilde{\a}}{k}\sum_{b,c}
%		f^{ba}_c\left( J^c(\phi)\overline{J}^b(\phi)-\overline{J}^c(\phi)J^b(\phi)\right)
%		\delta(\phi-\phi')\nn
%	&
	=(1-\widetilde{\a})\pd_\phi\left[\mathbf{J}^a(\phi)\delta(\phi-\phi')\right]
	-\frac{2\widetilde{\a}}{k}
	\sum_{b,c}f^{a}_{bc}\,\mathbf{P}^b(\phi)\mathbf{J}^c(\phi)
	\delta(\phi-\phi').
\ee
For the purely Abelian case where $f^a_{bc}=0$, the above algebra, together with \eqref{bms.al},
nicely describes an interpolation
between the Virasoro-Kac-Moody and BMS-Kac-Moody algebras. Note here, for $\a = 1$ the BMS-Kac-Moody structure in $\{\cJ(\a;\phi), \mathbf{P}^a(\phi')\}$ and $\{\cP(\a;\phi), \mathbf{P}^a(\phi')\}$ was also spotted in \cite{Hao:2021urq}, as the reader can notice the Fourier modes of $\mathbf{P}$ are just the $B_n$ modes as defined for null strings  \eqref{lmab}.
\medskip

The non-Abelian case still is bit problematic at the current level as it contains the presence of the last term in \eqref{calPbfP.alg} involving structure constants.
At this moment a clear understanding of the term  for non-Abelian case is missing. However
the term can be carefully treated once the system is quantized.
Naively, in the case of Dirac brackets, we can see this term involving structure constants gives rise to a contact term
\be
	f^{a}_{bc}\,\mathbf{P}^b(\phi)\mathbf{J}^c(\phi)
	=\frac{f^{a}_{bc}}{2} [\mathbf{P}^b(\phi),\mathbf{J}^c(\phi)]
	\sim f^{a}_{bc}f^{bc}_d\,\mathbf{P}^d(\phi)\delta(0)
\ee
where the divergent piece $\delta(0)$ should be removed by a suitable prescription of ordering of
operators as well as defining a vacua properly. Again, we have to note this is precisely the problem of ordering of $A,B$ oscillators for the null string theory case, which lead to the discovery of three distinct possible vacua \cite{Bagchi:2020fpr}.
However, since we are focusing only on classical case in the present work,
this point is beyond the scope at this moment. We plan to come back to this soon.

\subsubsection*{Hamiltonian and equations of motion}

Let us now elaborate more about our $\a$ dependent deformation, 
this time focusing on the Hamiltonian and the equations of motion.
The boosted Hamiltonian density in this theory is given by \footnote{We can be very much tempted to write down the analogue of a flow equation from here, which reads: $$\partial_{\widetilde{\a}}\widetilde{\cH}(\a;\phi)\sim J\overline{J}$$. }
\bea
	2\widetilde{\cH}(\a;\phi)
	&={T}_{\rm L}(\a;\phi)+{T}_{\rm R}(\a;\phi)
	\\
	&=\frac{1}{2\sqrt{1-\widetilde{\a}^2}}
	\left(\pd X^\mu \pd X_\mu+\overline{\pd}X^\mu\overline{\pd}X_\mu
	+2\widetilde{\a}\pd X^\mu\overline{\pd}X_\mu\right)
	\nn
	&=\sqrt{\frac{1+\widetilde{\a}}{1-\widetilde{\a}}}\dot{X}^2
	+\sqrt{\frac{1-\widetilde{\a}}{1+\widetilde{\a}}}{X'}^2
\ena
The conjugate momenta $\widetilde{\Pi}_\m$ can be calculated as
\be
	\widetilde{\Pi}_\m(\a;\phi)=\sqrt{\frac{1+\widetilde{\a}}{1-\widetilde{\a}}}\dot{X}_\m.
\ee
Now the Hamiltonian density reads:
\be
	\widetilde{\cH}(\a;\phi)=\sqrt{\frac{1-\widetilde{\a}}{1+\widetilde{\a}}}
	\left(\frac{\widetilde{\Pi}^2}{2}+\frac{{X'}^2}{2}\right)
\ee
Performing a Legendre transform via
$\pd\widetilde{\cH}/\pd\widetilde{\Pi}_\mu=\dot{X}_\m$,
the corresponding Lagrangian density is given by
\be\label{lagra}
	\widetilde{\cL}
	:=\widetilde{\Pi}_\mu\dot{X}^\mu-\widetilde{\cH}
	=\frac{1}{\kappa} \frac{\dot{X}^2}{2}-\kappa\frac{{X'}^2}{2}
\ee
with introducing a parameter
\be
	\kappa=\sqrt{\frac{1-\widetilde{\a}}{1+\widetilde{\a}}}=\frac{1-\a}{1+\a}
\ee
The equations of motion associated to this system reads
\be
	\left(\frac{1}{\kappa^2}\frac{\pd}{\pd \tau^2}-\frac{\pd^2}{\pd \phi^2}\right)X^\mu=0.
\ee
Comparing this with wave equation for the pullback field $X$, we can see $\kappa$ acts as a scaled ``speed of light".
Therefore, the limit $\a\to1$ corresponds to the limit where
the speed of light' $\kappa$ approaches to zero
(``ultra-relativistic'' limit), while $\a\to-1$ corresponds to the ``non-relativistic'' limit ($\kappa\to\infty$).
Under both of these limits the symmetries of the system boils down to the (classical part of) BMS$_3$ algebra due to the BMS/GCA duality in two dimensions \cite{Bagchi:2010zz}. This matches perfectly with what we discussed at the very beginning of this section. 

This duality can also be seen when one calculates the stress energy tensors from the Lagrangian density \eqref{lagra}. While the diagonal components $T^t_{~t} =-T^{\phi}_{~\phi} = \widetilde{\cH}$, making the stress tensor traceless, the off diagonal components are not equal anymore , in fact we can see,
\be
T^t_{~\phi}= \frac{1}{\kappa}\dot{X}X',~~~T^\phi_{~~t}= \kappa\dot{X}X'.
\ee
This means the limit $\a\to1$ corresponds to $T^\phi_{~~t}=0$, i.e. no energy flux is possible, signaling the emergence of a Carroll boost invariant theory \cite{deBoer:2021jej}, while $\a\to-1$ will give rise to $T^t_{~\phi}=0$, corresponding to a Galilei boost invariant theory. The seeming divergence in the other two components can actually be killed off by the simultaneous scaling of the coupling constant (tension) as in \eqref{URlim}.

Although the limit $\a\to-1$ ($\kappa\to\infty$) appears fine at the level of
equations of motion, it may seem pathological 
in a sense that it encounters a singularity
where the Hamiltonian density $\widetilde{\cH}$ diverges at the point.
To remove this singularity,
a useful procedure is to rescale the variables as
\be
	\kappa=\frac{1}{\kappa_D}
	\qquad\qquad
	X^\mu
	\ \longrightarrow \
	\sqrt{\kappa_{D}}X_{D}^\mu
	\qquad\qquad
	\widetilde{\Pi}^\mu
	\ \longrightarrow \
	\frac{1}{\sqrt{\kappa_D}}\widetilde{\Pi}_D^\mu
	\left(=\frac{1}{\sqrt{\kappa_D}}\dot{X}_D^\mu\right)
\ee
as well as $\,\widetilde{\cH}\,\longrightarrow \,\kappa_D^{-2}\widetilde{\cH}_D$,
so that the Hamiltonian density is replaced by the more suitable expression,
\be
	\widetilde{\cH}=\kappa\left(\frac{\widetilde{\Pi}^2}{2}+\frac{{X'}^2}{2}\right)
	\qquad\longrightarrow\qquad
	\widetilde{\cH}_{D}
	=\kappa_D^2\left(
	\frac{1}{\kappa_D^2}\frac{\widetilde{\Pi}_D^2}{2}+\frac{{X'}_D^2}{2}
	\right)
\ee
The proportionality constant between $\widetilde{\Pi}_D$ and $\dot{X}_D$ is defined so as to satisfy
${\pd \widetilde{\cH}_{D}}/{\pd \widetilde{\Pi}^\mu_D}=\dot{X}_{D}^{\mu}$.
The Lagrangian density correspondingly gets replaced by
\be
	\cL_D=\left.\widetilde{\Pi}_D\right._\mu\dot{X}_D^\mu-\widetilde{\cH}_D
	=\kappa_D\left(
	\frac{1}{\kappa_D}\frac{\dot{X}_D^2}{2}-\kappa_D\frac{{X'}_D^2}{2}\right).
\ee
The equations of motion now take again the form of the wave propagation equation
\be
	\left(\frac{1}{\kappa_D^2}\frac{\pd}{\pd \tau^2}-\frac{\pd^2}{\pd \phi^2}\right)X_D^\mu=0
\ee
though the ``speed of light'' is now given by $\kappa_D$, not $\kappa$.
Thus the ``non-relativistic'' limit of taking $\kappa\to\infty$ is equivalent to
the ``ultra-relativistic'' limit $\kappa_D\to0$ under the duality.
While in this case the limit $\kappa_D\to\infty$ ($\a\to1$) is pathological instead due to
 the divergence of the dual Hamiltonian density $\widetilde{\cH}_D$ at $\a=1$. This duality in ``speed of light" could be thought 
 of as equivalent to the treatise of \cite{Duval_2014} where two different notions of defining ``time'' has been considered 
 to measure propagation velocity of electromagnetic waves in Carroll or Galilei frames.

%\subsection*{Differential generators}

\bigskip

\section{``Asymmetrical boosts'' and null theories }\label{asyboost}
%Bosonic strings have been formulated in various fashions.
%Each formulations has its own advantages and disadvantages.
%With an intention of emphasizing which aspects,
%one may choose the most appropriate formulation
%according to their problems.
For bosonic string theory in flat space, the classical theory is usually studied by considering 
the Polyakov action in the conformal gauge \eqref{conaction}. Under the choice of conformal gauge,
the residual symmetries are two copies of the Virasoro algebra,
and important physical properties can be analyzed by applying well-developed techniques in $2d$ CFT.
\medskip

Although the conformal gauge is very well studied and offers symmetric structures,
other gauge choices have made their appearance in the literature as well.
For our purpose in this paper, we are also interested in some of the prominent other examples. Let us first talk about the one parameter family of gauges, called Hohm, Siegel and Zwiebach (HSZ) gauge \cite{abc},
which can be achieved by parametrizing the string worldsheet by a global transformation \footnote{One can note that compared to \eqref{Symm:z_LR}, this is a skewed boost transformation on the (anti)holomorphic coordinates. These can be compared to Galilean or Carrollian boosts, upto a scale factor, where either of the two coordinates remains absolute under these transformations.  }
\be \label{HSZ}
	\begin{pmatrix}
	z_{\rm L} \\ z_{\rm R}
	\end{pmatrix}
	=\frac{1}{\sqrt{1+\b}}
	\begin{pmatrix}
	1+\b &0 \\ -\b & 1
	\end{pmatrix}
	\begin{pmatrix}
	z \\ \overline{z}
	\end{pmatrix}
\ee
where $\b$ is some constant for our purposes, resulting in an action of form
\be
	\cL=\pd_{\rm L}X^\mu \pd_{\rm R}X_\mu=\pd X \bar{\pd} X+\b \left(\bar{\pd}X\right)^2
\ee
The resultant action no longer allows one to apply the conventional techniques of CFT in a straightforward manner, and instead leads to interesting results \cite{Siegel:2015axg}. As one can easily see, these asymmetric transformations on the (anti)-holomorphic coordinates cannot be called (quasi-) boosts in any way and they do not conserve the conformal gauge. For finite $\b$, the usual symmetries of the worldsheet are broken. The singular limit $\beta\to\infty$, better known as the HSZ limit, is where a special chiral theory appears.
\medskip

Taking a cue from the above discussion, we can introduce here a multi-parameter generalization of the worldsheet parametrization in the following form
\be
\label{z_LR}
	\begin{pmatrix}
	z_{\rm L} \\ z_{\rm R}
	\end{pmatrix}
	=\frac{1}{\sqrt{1+\b-\a\gamma}}
	\begin{pmatrix}
	1+\b & -\a \\ -\gamma & 1
	\end{pmatrix}
	\begin{pmatrix}
	z \\ \overline{z},
	\end{pmatrix}
\ee
where $\a,\b,\gamma$ are all numbers, and which includes the HSZ parameterisation as a special case.
Depending on the choice of parameters,
it will turn out that the resultant action
produces not only the HSZ gauge action, but also
other classes of interesting actions discussed in literature.
%Among them, on some isolated points
%the action achieves BMS symmetry, which has attracted attention recently
%in several contexts \rc{[Refs]}.
%It will also be discussed that
%one of the branches of parameter choice shall keep conformal
%invariance despite the fact that it is out of conformal gauge,
%and that, on its endpoints, the resultant action achieves BMS symmetry.
%

\subsection{Worldsheet actions}

For the class of parametrization on the string worldsheet \eqref{z_LR},
one may define the action
\be
	\cL
	=\pd_{\rm L}X^\mu \pd_{\rm R}X_\mu
	=\eta^{AB}\frac{\pd X^\mu}{\pd \xi^A}\frac{\pd X_\mu}{\pd \xi^B}
\ee
where the coordinates $\{\xi^A\}=\{z,\bar{z}\},\ \{\tau,\sigma\}$ 
($z=\tau+\sigma$ and $\bar{z}=\tau-\sigma$), correspond to the gauge choice where the worldsheet metric components are fixed to
\be
	\begin{pmatrix}
	\eta^{zz} & \eta^{z\bar{z}} \\ \eta^{\bar{z}z} & \eta^{\bar{z}\bar{z}}
	\end{pmatrix}
	=
	\begin{pmatrix}
	\frac{\a}{1+\b-\a\gamma}
	& \frac12\left(\frac{1+\b+\a\gamma}{1+\b-\a\gamma}\right) \\
	\frac12\left(\frac{1+\b+\a\gamma}{1+\b-\a\gamma}\right)
	& \frac{\gamma(1+\b)}{1+\b-\a\gamma}
	\end{pmatrix}
	\qquad
	\begin{pmatrix}
	\eta^{\tau\tau} & \eta^{\tau\sigma} \\ \eta^{\sigma\tau} & \eta^{\sigma\sigma}
	\end{pmatrix}
	=
	\begin{pmatrix}
	\frac{(1+\gamma)(1+\a+\b)}{4(1+\b-\a\gamma)}
	& \frac{\a-\gamma(1+\b)}{4(1+\b-\a\gamma)} \\
	\frac{\a-\gamma(1+\b)}{4(1+\b-\a\gamma)}
	& -\frac{(1-\gamma)(1-\a+\b)}{4(1+\b-\a\gamma)}
	\end{pmatrix}
\ee
Turning off all the parameters $\a=\b=\gamma=0$, the action coincides with the bosonic string action of Polyakov type with conformal gauge. Turning on $\b$ and $\gamma$ in such a way that $\b=\gamma$, while keeping $\a=0$, corresponds to the generic HSZ gauge. It should be emphasised that the parameters $\b$ and $\gamma$ can be chosen independently, and more crucial role is played by $\gamma$, while $\b$ serves as a normalization factor. It will be transparent by taking a look into the corresponding metric for the parameter choice
$(\a=0, \ \b, \ \gamma)$ \footnote{Note other choice of parameters, $(\a, \ \b, \ \gamma=0)$
is essentially same but merely flips
the roles of $z$ and $\bar{z}$.
}
\be\label{metrgam}
	\begin{pmatrix}
	\eta^{zz} & \eta^{z\bar{z}} \\ \eta^{\bar{z}z} & \eta^{\bar{z}\bar{z}}
	\end{pmatrix}
	=
	\begin{pmatrix}
	0 & \frac12 \\
	\frac12 & \gamma
	\end{pmatrix}
	\qquad\quad
	\begin{pmatrix}
	\eta^{\tau\tau} & \eta^{\tau\sigma} \\ \eta^{\sigma\tau} & \eta^{\sigma\sigma}
	\end{pmatrix}
	=
	\begin{pmatrix}
	\frac{1+\gamma}{4}
	& -\frac{\gamma}{4} \\
	-\frac{\gamma}{4}
	& -\frac{1-\gamma}{4}
	\end{pmatrix}
\ee
where the parameter $\b$ no longer appears in the metric.
The metric on $(\tau,\sigma)$-coordinates clearly shows
the presence of off-diagonal components once we turn on non-zero $\gamma$,
resulting in a metric non-compliant with conformal gauge, indicating that
the conventional techniques of CFT cannot be applied to the system.
There are two interesting points to be mentioned, first the case of $\gamma=\pm1$ where one of the
diagonal components vanish. The Lagrangian on these points take the form:
\be
	\cL_{\gamma=\pm1}=
	\begin{cases}
	\ \frac12 \dot{X}\left(\dot{X}-X'\right) \qquad(\gamma=1)\\
	\ \frac12 {X}'\left(\dot{X}-X'\right) \qquad(\gamma=-1).
	\end{cases}
\ee
These are identical to the action (or a space-time flip thereof) known as Floreanini-Jackiw Lagrangian for a chiral boson \cite{Floreanini:1987as}. One can show from a Noether procedure that on-shell the symmetries of these actions boil down to one copy of the Virasoro algebra \cite{Bagchi:2022eav}.

Another interesting point in the parameter space is obtained by taking the limit $\gamma\to\infty$, where the worldsheet metric \eqref{metrgam} becomes degenerate: a telltale sign of a null string limit.
Normalizing the action appropriately, the resulting action takes the form
\be\label{ambit}
	\widehat{\cL}_{\gamma\to\infty}=\left(\bar{\pd}X\right)^2.
\ee
which coincides with the action of Ambitwistor string  \cite{Casali:2016atr, Casali:2017zkz}. This choice is a singular gauge, where all the dependence on the $z$ direction drops out. Intriguingly, one can check that this action can also be obtained from the second order ILST action \eqref{LST} using the gauge choice $V^\a = (1,-1)$. It has already been shown  \cite{Casali:2016atr} that the Ambitwistor action is classically BMS invariant on shell, as we would expect. 

\subsection{On-shell symmetries for the null case}
We now discuss the fate of symmetries for these theories. Since our focus is on null limits of these ``boosted'' theories, the class of action corresponding to \eqref{ambit} is a good choice to focus on. As we can also see, we can generate these classes of action via a $\b\to \infty$ limit on the HSZ parameterisation \eqref{HSZ}, where again only one chiral sector survives in the action.
Let us start with the transformation of currents corresponding to  \eqref{HSZ}:
\be
	\begin{pmatrix}
	J^a_{\rm L} \\ J^a_{\rm R}
	\end{pmatrix}
	=\frac{1}{\sqrt{1+\b}}
	\begin{pmatrix}
	1 & \b \\ 0 & 1+\b
	\end{pmatrix}
	\begin{pmatrix}
	J^a(z) \\ \overline{J}^a(\bar{z})
	\end{pmatrix}
\ee
and write the suitably $\b$ dependent symmetry generators as deformed versions of \eqref{UndeformJP}
\be
\begin{aligned}\label{bmsf}
	\cJ(\b;\phi)
	&=\frac{1}{k\b}\sum_a \left(J^a_{\rm L}(\phi)J^a_{\rm L}(\phi)-J^a_{\rm R}(\phi)J^a_{\rm R}(\phi)\right)\\
	&=\frac{1}{k(1+\b)}\sum_a \Big[ J^a(\phi){J}^a(\phi) - \overline{J}^a(\phi)\overline{J}^a(\phi) -2\b\left(\overline{J}^a(\phi)\overline{J}^a(\phi)-J^a(\phi)\overline{J}^a(\phi)\right)  \Big]\\
	\cP(\b;\phi)
	&=\frac{1}{k\b}\sum_a \left(J^a_{\rm L}(\phi)J^a_{\rm L}(\phi)+J^a_{\rm R}(\phi)J^a_{\rm R}(\phi)\right)\\
	&=
	\frac{1}{k\b(1+\b)}\sum_a \Big[ J^a(\phi){J}^a(\phi) + \overline{J}^a(\phi)\overline{J}^a(\phi) +2\b(1+\b)\overline{J}^a(\phi)\overline{J}^a(\phi)+2\b J^a(\phi)\overline{J}^a(\phi)  \Big].
\end{aligned}
\ee
 It is difficult to see the symmetries in this general form of the generators, but taking a consistent $\b\to \infty$ limit and a simple rescaling, we can get the $\b$ independent generators:
\be
\cJ_{\infty} = -\overline{T}+ \frac{1}{k}\sum_aJ^a(\phi)\overline{J}^a(\phi),~~~\cP_{\infty} = \overline{T},
\ee
so that there are only two independent skewed set of generators $ \frac{1}{k}\sum_aJ^a(\phi)\overline{J}^a(\phi)$ and $\overline{T}$.
To find the on-shell symmetries in this limit, the simplest way is to resort to the equations of motion and mode expansion corresponding to \eqref{ambit}: 
\be{}
\bar{\pd}^2 X=0,~~~X(z,\overline{z}) = \A(z) + (\overline{z}+z)\B(z)
\ee
The form of the solution makes sure of periodicity on the infinite cylinder $\phi \sim \phi+2\pi$.
We assume $\A,\B$ are arbitrary functions with Fourier expansions
\be{}
\A(z) = \sum_{n}\A_n e^{inz},~~\B(z) = \sum_{n}\B_n e^{inz}.
\ee
The equal-time Poisson brackets for the modes taking the following form:
\be{}
\{\A_n, \B_m\}_{\text{PB}} = \delta_{n+m,0},~~\{\A_n, \A_m\}_{\text{PB}} = \{\B_n, \B_m\}_{\text{PB}}  = 0.
\ee
A simple calculation yields the symmetry generators in terms of Fourier modes of $\cJ, \cP$:
\begin{eqnarray}
\cJ_{n\infty} &=& \sum_{m}\Big( im\A_{-m}\B_{n+m} - in(\overline{z}+z)\B_{-m}\B_{n+m} \Big)= L_n - in\t M_n\\
\cP_{n\infty} &=& \sum_{m} \B_{-m}\B_{n+m} = M_n.
\end{eqnarray}
These modes again satisfy the Poisson brackets corresponding to the BMS$_3$ algebra \eqref{bms1}. 

\bigskip

\section{Discussions and Conclusions}
\subsection*{Summary}
In this paper, we discussed a mechanism to ``flow" from a $2d$ CFT to a $2d$ BMSFT (or GCFT) by explicitly boosting the theory, which mirrors earlier revelations for tensionless limits on tensile strings, but unlike the latter, does not impose a direct contraction of the coordinates. We ``boosted'' the $2d$ CFT quantities written in (anti)holomorphic form by a special linear transformation that turns out to be equivalent to the contraction procedure. For a symmetrically defined transformation on both Left and Right sector of the theory, we found this transformation effectively generates a $J\overline{J}$-like deformation term in the CFT Hamiltonian, whose strength is proportional to the boost parameter. As we go to the extreme case where the boost transformation becomes non-invertible, the CFT quantities undergo a ``phase" transition, and BMS invariance sets in. We also could explicitly see this in the symmetry algebra which, written in proper basis, smoothly transitions from two copies of Virasoro algebra to the BMS$_3$ algebra when the boost parameter touches the extreme allowed values. 
\medskip

This simple way to traverse from one theory to another in the space of symmetries also turned out to give interesting results when left and right sectors of the conformal theory were not transformed in the same footing. In a later section we showed, ``asymmetric boosts", which break explicit Lorentz invariance, leads one from CFT$_2$ to certain intriguing chiral theories. We focussed on certain special points in this parameter space, corresponding to theories well known in literature.  Especially one could formulate a degenerate point, where the intrinsic metric of the theory becomes null, and as expected from physical considerations, emergent BMS$_3$ invariance (on-shell) can be found.
\subsection*{Boost vs Acceleration}

As far as this work is concerned, we have stuck to effectively boosting our CFT coordinates to show a transition into BMS symmetries. This makes concrete the lore in the literature that BMSFTs are infinitely boosted versions of CFTs. However, especially in the arena of tensionless strings, it has been shown over the last couple years, that a similar transition happens when one \textit{``accelerates''} the relativistic conformally invariant worldsheet \cite{Bagchi:2020ats, Bagchi:2021ban}. While a finitely accelerated $2d$ theory maintains two copies of Virasoro as classical symmetry algebra, the associated vacuum structure changes with acceleration due to an analogue of the Unruh effect. BMS$_3$ then emerges at the limit of infinite acceleration, or to go by the interpretation of \cite{Bagchi:2021ban}, when a relativistic string worldsheet falls into a black hole horizon. This is effectively a quantum mechanical statement following from \eqref{CC} and \eqref{ABa}, which, when combined gives a relationship between $C$ (Carrollian) oscillators and $\alpha$ (Relativistic) oscillators:
 \begin{align}
C^{\mu}_{n}&=\frac{1}{2}\Big(\sqrt{\epsilon}+\frac{1}{\sqrt{\epsilon}}\Big)\alpha^{\mu}_{n}+\frac{1}{2}\Big(\sqrt{\epsilon}-\frac{1}{\sqrt{\epsilon}}\Big)\Tilde{\alpha}^{\mu}_{-n} \nonumber \\
\Tilde{C}^{\mu}_{n}&=\frac{1}{2}\Big(\sqrt{\epsilon}-\frac{1}{\sqrt{\epsilon}}\Big)\alpha^{\mu}_{-n}+\frac{1}{2}\Big(\sqrt{\epsilon}+\frac{1}{\sqrt{\epsilon}}\Big)\Tilde{\alpha}^{\mu}_{n},
\label{infvel}
\end{align}
which is a Bogoliubov transformation that keeps the structure of canonical commutation relations unchanged for all possible values of $\e$.\footnote{Although this relation has been written down for $\e\to 0$, one can extrapolate this to realize a connected transformation, at least near $\e = 1$, where $C$ and $\a$ oscillators become the same.} At the quantum level this points to a one parameter continuous evolution of vacua, which takes the CFT one to the BMS one. 

Now we can ask the question, whether we can see a similar story in our formulation as well. Our initial investigations seem to suggest that this is indeed possible and the clue lies in the observation that the structure of  \eqref{currentchange} can be read off as 
\begin{eqnarray}
J^a_{L } = \cosh\theta~J^a- \sinh\theta~\overline{J}^a, \quad  J^a_{R } = -\sinh\theta~J^a +\cosh\theta~\overline{J}^a.
\end{eqnarray}
Here $\cosh\theta = \frac{1}{\sqrt{1-\a^2}}$ and $\sinh\theta =- \frac{\a}{\sqrt{1-\a^2}}$ can be interpreted as Bogoliubov coefficients. We hope to report on to this interesting problem soon. 

\subsection*{Future directions}
This work was aimed at introducing a straightforward formalism of boosting a $2d$ CFT to a $2d$ BMSFT, and that out of the way, there are clear things we aim to look at in the near future. First of all, one should note that this work has been presented in a way suitable for a $2d$ CFT (or a string worldsheet theory). However, there must be a generalization of this formulation into higher dimensions, where one doesn't have the luxuries associated to $2d$ and the CCA-GCA isomorphism. Since the kinematical structure of Carrollian algebras remain the same in all dimensions, and the contraction from relativistic theories are no different as well, one should be able to understand them in the ``boost'' perspective as presented in this work. This might be more challenging while dealing with more complicated nature of transformations, but should lead one to insightful details about these theories. 
\medskip

That being said, there are a plethora of problems just in $2d$ which one can hope to tackle with our formalism. The very first thing that comes into mind is to quantify Carrollian spinors. Since the speed of light changes gradually in our ``boosts'', that would indeed mean Clifford algebras would also flow from that of a relativistic one to a degenerate BMS cousin of it. This would mean structure of Dirac matrices will also be dependent on the boost parameter. This has obvious consequences, for example in the null worldsheet theory with Super-BMS algebra as residual symmetries. It has been shown that the Carrollian cousin of $\mathcal{N}=1$ supersymmetric string has two distinct versions of the Super-BMS algebra as worldsheet gauge symmetries, based on how the spinors are treated under contraction. These two point out two distinct and inequivalent representations of the degenerate Clifford algebra (called homogeneous and inhomogeneous). One might envision these two as two separate fixed points in the space of ``boosted'' spinors, but this obviously has to be established mathematically. 
\medskip

 Lastly, we were content with the classical version of the symmetry algebras we used throughout this work. It would be nice to instead work with a full relativistic Wess-Zumino-Witten model and describe the infinitely boosted version following methods described in this work. In that case we should be able to see how physical states flow under these transformations, and e.g. how the central charges change as we approach the BMS invariant point. In a broad sense both Abelian and Non-Abelian WZW models can be addressed in this regard, which would be much more general than our particular focus on worldsheet string theories in this paper. It would also be interesting to see how modular transformations and partition functions of relativistic theories get deformed under such infinite boosts. We hope to discuss these aspects in future work.  

\bigskip

\section*{Acknowledgements}
We thank Pulastya Parekh for initial discussions. 

\medskip

\noindent AB is partially supported by a Swarnajayanti Fellowship of the Department of Science and Technology and Science and Engineering Research Board (SERB) and also by the following SERB grants SB/SJF/2019-20/08, CRG/2020/002035. The work of ArB is supported by the Quantum Gravity Unit of the Okinawa Institute of Science and Technology Graduate University (OIST). ArB would like to thank Kyoto University and IIT Kanpur for kind hospitality during the last stages of this work. The work of HM is supported by the Institute for Basic Science (IBS-R003-D1).

\newpage 
\appendix 

%%%%%%%%%%%%%%%%%%%%%%%%%%%%%%%%%%%%%%%%%%%
\section{Coordinate transforms}\label{apa}

%%%%%%%%%%%%%%%%%%%%%%%%%%%%%%%%%%%%%%%%%%%
\subsection{Lorentz boosts}
Let us first recall the notion of a Lorentz boost,
leaving the flat Minkowski metric $\eta_{ab}={\rm diag}(1,-1)$ invariant.
A general form of proper Lorentz boost (connected component of unity)
is given by
\be\label{Lorentz_boost-matrices}
	\begin{pmatrix}
	\tau' \\ \sigma'
	\end{pmatrix}
	=
	\begin{pmatrix}
	\cosh\phi & \sinh\phi \\ \sinh\phi & \cosh\phi
	\end{pmatrix}
	\begin{pmatrix}
	\tau \\ \sigma
	\end{pmatrix}
\ee
Introducing LC coordinates ($z'=\tau'+\sigma'$ and $\bar{z}'=\tau'-\sigma'$)
\be
	\begin{pmatrix}
	z' \\ \bar{z}'
	\end{pmatrix}
	=
	L_\phi
	\begin{pmatrix}
	z \\ \bar{z}
	\end{pmatrix}
	\qquad\qquad
	L_\phi
	=
	\begin{pmatrix}
	e^{\phi} & 0 \\ 0 & e^{-\phi}
	\end{pmatrix}
\ee
The relation can be summarized by the following diagram.
\be
\begin{array}{ccc}
	(\tau,\sigma) & \overset{\text{Lorentz boost}}{\longrightarrow} & (\tau',\sigma') \\
	\downarrow & & \downarrow \\ 
	(z,\bar{z}) & \overset{L}{\longrightarrow} & (z',\bar{z}')
\end{array}
\ee
Any proper Lorentz boosts are represented on LC coordinates
by a class of matrices of form
\be
	\mathbb{L}_{\rm LC}
	=\left\{L=\begin{pmatrix} \Lambda & 0 \\ 0 & \Lambda^{-1}\end{pmatrix}
	\left.\right|\,\Lambda \geq0\right\}
\ee

%%%%%%%%%%%%%%%%%%%%%%%%%%%%%%%%%%%%%%%%%%%
\subsection{Boosts beyond Lorentz}
Let us relax the conditions so that the resultant metric under said transformation is invariant up-to a scaling of the coordinates,
i.e. $\tilde{\eta}_{ab}={\rm diag}(a^2,-b^2)$ with $(a,b\geq0)$. We can always use the conformal symmetries in $2d$ to rescale such a metric to the Minkowski one, and this does not break the structure of the conformal gauge.
The representation matrix of such transformations
can be figured out as follows.
\be
	\begin{pmatrix}
	\tau \\ \sigma
	\end{pmatrix}
	=
	\begin{pmatrix}
	A & B \\ C & D
	\end{pmatrix}
	\begin{pmatrix}
	\tilde{\tau} \\ \tilde{\sigma}
	\end{pmatrix}
\ee
where $(\tau,\sigma)$ is assumed to be defined on Minkowski metric and then
\be
	\left(d\tau\right)^2-\left(d\sigma\right)^2
	=\left(A^2-C^2\right)\left(d\tilde{\tau}\right)^2
	-\left(D^2-B^2\right)\left(d\tilde{\sigma}\right)^2
	+2\left(AB-CD\right)d\tilde{\tau}d\tilde{\sigma}
\ee
The condition for new metric to be of form $\tilde{\eta}_{ab}={\rm diag}(a^2,-b^2)$ imposes
\be
	\begin{cases}
	\ A^2-C^2=a^2\\
	\ D^2-B^2=b^2\\
	\ AB-CD=0
	\end{cases}
\ee
The first and second relations are solved by parameterizing
\be
	A=a\,\cosh\phi_a
	\qquad
	B=b\,\sinh\phi_b
	\qquad
	C=a\,\sinh\phi_a
	\qquad
	D=b\,\cosh\phi_b
\ee
and the third relation imposes $\phi_a=\phi_b=\tilde{\phi}$.
For this case, since 
$\left(d\tau\right)^2-\left(d\sigma\right)^2=
a^2\left(d\tilde{\tau}\right)^2-b^2\left(d\tilde{\sigma}\right)^2$,
one may introduce analogues of ``LC coordinates'' by
\be
	\tilde{z}=a\,\tilde{\tau}+b\,\tilde{\sigma}
	\qquad\quad
	\bar{\tilde{z}}=a\,\tilde{\tau}-b\,\tilde{\sigma}
\ee
Then the representation matrix on ``LC coordinates''  for the transform is
\be
	\begin{pmatrix}
	\tilde{z} \\ \bar{\tilde{z}}
	\end{pmatrix}
	=\begin{pmatrix}
	e^{\tilde{\phi}} & 0 \\
	0 & e^{-\tilde{\phi}}
	\end{pmatrix}
	\begin{pmatrix}
	z \\ \bar{z}
	\end{pmatrix}
\ee
The relation can be summarized by the following diagram.
\be
\begin{array}{ccc}
	(\tau,\sigma) & \overset{\text{``boost''}}{\longrightarrow} & (\tilde{\tau},\tilde{\sigma}) \\
	\downarrow & & \downarrow \\ 
	(z,\bar{z}) & \overset{\tilde{L}}{\longrightarrow} & (\tilde{z},\bar{\tilde{z}})
\end{array}
\ee
The ``boosts'' are represented on ``LC coordinates''
by a class of matrices of form
\be
	\tilde{\mathbb{L}}_{\text{``LC''}}
	=\left\{\tilde{L}=\begin{pmatrix} \Lambda & 0 \\ 0 & \Lambda^{-1}\end{pmatrix}
	\left.\right|\,\Lambda \geq0\right\}
\ee
The separation of variables takes place in the same manner as
the (anti-)holomorphic sectors 
(or left-/right-movers) of closed strings under the gauge choice
of diagonal metric.

While introducing another pair of coordinates,
\be
	{z}_{\rm L}=\tilde{\tau}+\tilde{\sigma}
	\qquad\quad
	{z}_{\rm R}=\tilde{\tau}-\tilde{\sigma}
\ee
in a similar manner, one finds our notion of ``boost'' is represented by
\be
	\begin{pmatrix}
	{z}_{\rm L} \\ {z}_{\rm R}
	\end{pmatrix}
	=\frac{1}{ab}\begin{pmatrix}
	\frac{a+b}{2}\,e^{-\tilde{\phi}} & -\frac{a-b}{2}\,e^{\tilde{\phi}} \\
	-\frac{a-b}{2}\,e^{-\tilde{\phi}} & \frac{a+b}{2}\,e^{\tilde{\phi}}
	\end{pmatrix}
	\begin{pmatrix}
	z \\ \bar{z}
	\end{pmatrix}
\ee
The one we have called ``symmetric boosts'' \eqref{Symm:z_LR} in the main text,
is the special case when $\tilde{\phi}=0$  and $a=b^{-1}=\sqrt{\frac{1+\a}{1-\a}}$,
which can also be parametrized by $a=b^{-1}=e^{\phi}$, giving rise to:
\be
	\begin{pmatrix}
	{z}_{\rm L} \\ {z}_{\rm R}
	\end{pmatrix}
	=
	\begin{pmatrix}
	\cosh\phi & \sinh\phi \\ \sinh\phi & \cosh\phi
	\end{pmatrix}
	\begin{pmatrix}
	z \\ \bar{z}
	\end{pmatrix}
\ee
resulting in the same form as \eqref{Lorentz_boost-matrices} but the vectors
on which the matrices act are replaced by $(\tau,\sigma)\to(z,\bar{z})$.
Interestingly, from this viewpoint, the global boost \eqref{HSZ} achieving the HSZ gauge
can be regarded essentially as (normalized version of) Galilei transforms
\be
	\begin{pmatrix}
	\tau' \\ \sigma'
	\end{pmatrix}
	=
	\begin{pmatrix}
	1 & 0 \\ -\b & 1
	\end{pmatrix}
	\begin{pmatrix}
	\tau \\ \sigma
	\end{pmatrix}
\ee
but replacing formally the coordinates $(\tau,\sigma)\to(z,\bar{z})$. 
These observations explain why transformations \eqref{Symm:z_LR}
and \eqref{HSZ} are something special,
owing to the fact that the two sets of matrices corresponding to those transformations
respectively form one-parameter subgroups in $\rm{SL}(2,\mathbb{R})$,
along which the speed of light varies:
\be
\begin{aligned}
	\mathbb{M}_{\text{``boosts''}}
	&=\left\{\left.
	\begin{pmatrix}
	\cosh\phi & \sinh\phi \\ \sinh\phi & \cosh\phi
	\end{pmatrix}\right|
	\phi\in(-\infty,+\infty)
	\right\}
	\subset \rm{SL}(2,\mathbb{R})
	\\
	\mathbb{M}_{\text{HSZ}}
	&=\left\{\left.
	\frac{1}{\sqrt{1+\b}}
	\begin{pmatrix}
	1+\b & 0 \\ -\b & 1
	\end{pmatrix}\right|
	\b\in(-1,+\infty)
	\right\}
	\subset \rm{SL}(2,\mathbb{R})
\end{aligned}
\ee

\section{Explicit calculations}\label{apb}
\subsection{Current algebra brackets}

The detailed calculations to prove the relations are as follows.
Noting the point-splitting prescription\footnote{
The differential arose from the Poisson bracket acts on the other fields
possessing the same argument, and the identification of the arguments by virtue of
the delta function can be applied only after all the differentials evaluated.},
\bea
	\{T(\phi),J^b(\phi')\}_{\rm PB}
	&=\frac{2}{k}\sum_a\{J^a(\phi),J^b(\phi')\}_{\rm PB}J^a(\phi)\nn
	&=\frac{2}{k}\sum_a
	\left(f^{ab}_c J^c(\phi)+\frac{k}{2}\delta^{ab}\pd_\phi\right)\delta(\phi-\phi')J^a(\phi)\nn
	&=\pd_\phi\left(\delta(\phi-\phi')J^b(\phi)\right)\nn
	&=J^b(\phi)\pd_\phi\delta(\phi-\phi')+\pd_\phi J^b(\phi)\delta(\phi-\phi')
\ena
and
\bea
	\{T(\phi),T(\phi')\}_{\rm PB}
	&=\frac{4}{k^2}\sum_{a,b}\{J^a(\phi),J^b(\phi')\}_{\rm PB}J^a(\phi)J^b(\phi')\nn
	&=\frac{4}{k^2}\sum_{a,b}
	\left(f^{ab}_c J^c(\phi)+\frac{k}{2}\delta^{ab}\pd_\phi\right)\delta(\phi-\phi')
	J^a(\phi)J^b(\phi')\nn
	&=\frac{2}{k}\sum_{a}
	\pd_\phi \left(\delta(\phi-\phi')
	J^a(\phi)J^a(\phi')\right)\nn
	&=\frac{2}{k}\sum_{a}
	\left(J^a(\phi)J^a(\phi')\pd_\phi\delta(\phi-\phi')
	+\pd_\phi J^a(\phi)J^a(\phi')\delta(\phi-\phi')
	\right)\nn
	&=2T(\phi)\pd_\phi\delta(\phi-\phi')
	+\pd_\phi T(\phi)\delta(\phi-\phi')
\ena

For the choice of $\overline{J}^a(\phi)=\sum_m \overline{J}_m^a e^{+im\phi}$
where the corresponding Poisson bracket is
\be
	\{\overline{J}^a(\phi), \overline{J}^b(\phi')\}_{\rm PB}
	=\left(f^{ab}_c \overline{J}^c(\phi)-\frac{k}{2}\delta^{ab}\pd_\phi\right)\delta(\phi-\phi')
\ee
the corresponding energy momentum tensor shall be
\be
	\overline{T}(\phi)=\frac{1}{k}\sum_a \overline{J}^a(\phi)\overline{J}^a(\phi)
\ee
so that in their algebraic relations signs are flipped as follows
\bea
	\{\overline{T}(\phi),\overline{J}^a(\phi')\}_{\rm PB}
	&=-\overline{J}^a(\phi)\pd_\phi\delta(\phi-\phi')-\pd_\phi \overline{J}^a(\phi)\delta(\phi-\phi')
	\\
	\{\overline{T}(\phi),\overline{T}(\phi')\}_{\rm PB}
	&=-2\overline{T}(\phi)\pd_\phi\delta(\phi-\phi')
	-\pd_\phi \overline{T}(\phi)\delta(\phi-\phi')
\ena
\subsection{BMS brackets}\label{appc}
Keeping the point-splitting prescription in mind, one finds,
as $\cJ=T-\overline{T}$ and
$\cP=T+\overline{T}+(2\widetilde{\a}/k)\sum_a J^a\bar{J}^a$, 
\bea
	\{\cJ(\phi),\cP(\phi')\}_{\rm PB}
	&=
	\{T(\phi),T(\phi')\}_{\rm PB}
	-\{\overline{T}(\phi),\overline{T}(\phi')\}_{\rm PB}
	\nn&\qquad
	+\frac{2\widetilde{\a}}{k}\{T(\phi),J^a(\phi')\}_{\rm PB}
		\overline{J}^a(\phi')
	-\frac{2\widetilde{\a}}{k}\{\overline{T}(\phi),\overline{J}^a(\phi')\}_{\rm PB}
		J^a(\phi')
	\nn&
	=2T(\phi)\pd_\phi\delta(\phi-\phi')
	+\pd_\phi T(\phi)\delta(\phi-\phi')
	\nn&\qquad
	+2\overline{T}(\phi)\pd_\phi\delta(\phi-\phi')
	+\pd_\phi \overline{T}(\phi)\delta(\phi-\phi')\nn
	&\qquad +\frac{2\widetilde{\a}}{k}
	J^a(\phi)\overline{J}^a(\phi')\pd_\phi\delta(\phi-\phi')
	+\frac{2\widetilde{\a}}{k}
	\left(\pd_\phi J^a(\phi)\right)\overline{J}^a(\phi')\delta(\phi-\phi')\nn
	&\qquad +\frac{2\widetilde{\a}}{k}
	J^a(\phi)\overline{J}^a(\phi')\pd_\phi\delta(\phi-\phi')
	+\frac{2\widetilde{\a}}{k}
	J^a(\phi)\left(\pd_\phi \overline{J}^a(\phi')\right)\delta(\phi-\phi')\nn
	&=\{\pd_\phi,\cP(\phi)\}\delta(\phi-\phi')
\ena
for which the value of $\widetilde{\a}$ is not significant.
While $\{\cP,\cP\}_{\rm PB}$ is calculated as follows
\bea
	\{\cP(\phi),\cP(\phi')\}_{\rm PB}
	&=
	\{T(\phi),T(\phi')\}_{\rm PB}
	+\{\overline{T}(\phi),\overline{T}(\phi')\}_{\rm PB}
	\nn&\qquad
	+\frac{2\widetilde{\a}}{k}\{T(\phi),J^a(\phi')\}_{\rm PB}
		\overline{J}^a(\phi')
	+\frac{2\widetilde{\a}}{k}\{\overline{T}(\phi),\overline{J}^a(\phi')\}_{\rm PB}
		J^a(\phi')
	\nn&\qquad
	+\frac{2\widetilde{\a}}{k}\{J^a({\phi}),T(\phi')\}_{\rm PB}
		\overline{J}^a({\phi})
	+\frac{2\widetilde{\a}}{k}\{\overline{J}^a({\phi}),\overline{T}(\phi')\}_{\rm PB}
		J^a({\phi})
	\nn&\qquad
	+\frac{4\widetilde{\a}^2}{k^2}
	\{J^a({\phi})\overline{J}^a({\phi}),J^b(\phi')\overline{J}^b(\phi')\}_{\rm PB}
\ena
where the second and the third lines cancel with each other,
whereas the final line is
\be
	\{J^a({\phi})\overline{J}^a({\phi}),J^b(\phi')\overline{J}^b(\phi')\}_{\rm PB}
	=-\frac{k^2}{4}\left(\{\overline{T}(\phi),\overline{T}(\phi')\}_{\rm PB}
	+\{T(\phi),T(\phi')\}_{\rm PB}
	\right)
\ee

thus in total
\be
	\{\cP(\phi),\cP(\phi')\}_{\rm PB}
	=(1-\widetilde{\a}^2)\{\partial_{\phi},\cJ(\phi)\}\delta(\phi-\phi')
\ee

\newpage

\bibliographystyle{JHEP}
\bibliography{ref}

\end{document}